# Compton-Thick AGN in the NuSTAR era VI: The observed Compton-thick fraction in the Local Universe

N. Torres-Albà,[1] S. Marchesi,[1,2] X. Zhao,[1] M. Ajello,[1] R. Silver,[1] T. T. Ananna,[3] M. Baloković,[4,5] P.B. Boorman,[6,7] A. Comastri,[2] R. Gilli,[2] G. Lanzuisi,[2] K. Murphy,[8] C. M. Urry,[4,9] and C. Vignali[10,2]

[1] *Department of Physics and Astronomy, Clemson University, Kinard Lab of Physics, Clemson, SC 29634, USA*
[2] *INAF - Osservatorio di Astrofisica e Scienza dello Spazio di Bologna, Via Piero Gobetti, 93/3, 40129, Bologna, Italy*
[3] *Department of Physics & Astronomy, Dartmouth College, 6127 Wilder Laboratory, Hanover, NH 03755, USA*
[4] *Yale Center for Astronomy & Astrophysics, 52 Hillhouse Avenue, New Haven, CT 06511, USA*
[5] *Department of Physics, Yale University, P.O. Box 208120, New Haven, CT 06520, USA*
[6] *Astronomical Institute, Academy of Sciences, Boční II 1401, CZ-14100 Prague, Czech Republic*
[7] *Department of Physics & Astronomy, Faculty of Physical Sciences and Engineering, University of Southampton, Southampton SO17 1BJ, UK*
[8] *Department of Physics, Skidmore College, Saratoga Springs, NY 12866, USA*
[9] *Department of Physics, Yale University, P.O. Box 2018120, New Haven, CT 06520, USA*
[10] *Dipartimento di Fisica e Astronomia, Università degli Studi di Bologna, via Gobetti 93/2, I-40129 Bologna, Italy*

## ABSTRACT

We present the analysis of simultaneous *NuSTAR* and *XMM-Newton* data of 8 Compton-thick (CT-) active galactic nuclei (AGN) candidates selected in the *Swift*-Burst Alert Telescope (BAT) 100 month survey. This work is part of an ongoing effort to find and characterize all CT-AGN in the local ($z \leq 0.05$) Universe. We used two physically motivated models, `MYTorus` and `borus02`, to characterize the sources in the sample, finding 5 of them to be confirmed CT-AGN. These results represent an increase of $\sim 19\%$ over the previous *NuSTAR*-confirmed, BAT-selected CT-AGN at $z \leq 0.05$, bringing the total number to 32. This corresponds to an observed fraction of $\sim 8\%$ of all AGN within this volume-limited sample, although it increases to $20 \pm 5\%$ when limiting the sample to $z \leq 0.01$. Out of a sample of 48 CT-AGN candidates, selected using BAT and soft ($0.3-10$ keV) X-ray data, only 24 are confirmed as CT-AGN with the addition of the *NuSTAR* data. This highlights the importance of *NuSTAR* when classifying local obscured AGN. We also note that most of the sources in our full sample of 48 Seyfert 2 galaxies with *NuSTAR* data have significantly different line-of-sight and average torus column densities, favouring a patchy torus scenario.

## 1. INTRODUCTION

Active galactic nuclei (AGN) are accreting supermassive black holes in the central regions of galaxies. AGN are mainly believed to be responsible for the cosmic X-ray background (CXB): the diffuse X-ray emission from a few keV to a few hundred keV (e.g. Marshall et al. 1980; Comastri et al. 1995; Alexander et al. 2003; Gilli et al. 2007; Ueda et al. 2014). In particular, at the peak of the CXB (20−30 keV; Ajello et al. 2008), a significant fraction of the emission ($\sim 15 - 20\%$; Gilli et al. 2007; Ananna et al. 2019) is attributed to a large population of Compton-thick AGN (CT-AGN, sources with obscuring hydrogen column densities $N_\mathrm{H} \geq 10^{24}$ cm$^{-2}$).

Nonetheless, in the nearby Universe ($z \leq 0.1$) the observed fraction of X-ray selected CT-AGN with respect to the total population is between 5 and 10% (e.g. Vasudevan et al. 2013; Ricci et al. 2015). This is much lower than the fractions predicted by most AGN population synthesis models which require, to properly explain the CXB, values between 20% (Ueda et al. 2014) and 50% (Ananna et al. 2019). CT-AGN are difficult to detect in X-rays due to the large obscuration in their line-of-sight, which significantly suppress their intrinsic emission, particularly at energies below 10 keV. Taking this into account, observational bias estimates recover a fraction of CT-AGN in the local Universe of at least $\sim 20\%$ (e.g. Brightman & Nandra (2011); Ricci et al. (2015); and refer to Sect. 4.2 of Burlon et al. (2011) for details on underlying model assumptions).

Infrared surveys obtain high fractions of Compton-thick AGN, based on X-ray non-detections of infrared-selected objects (e.g. Stern et al. 2014; Del Moro et al. 2016; Asmus et al. 2020). Without an X-ray characterization, however, it is difficult to estimate their contribution to the CXB. The same is true for any large optically or infrared-selected samples, which generally recover larger CT-AGN fractions, compared to X-ray studies. It is thus still necessary to reconcile the observed

X-ray-selected CT-AGN fraction with CXB model predictions.

In order to successfully detect and characterize these heavily obscured sources, it is necessary to use hard (> 10 keV) X-ray observatories, particularly in the local Universe[1]. The Burst Alert Telescope (BAT) on board the Neil Gehrels Swift (hereafter *Swift*) observatory is an instrument designed to provide critical gamma-ray burst triggers as it surveys the whole sky (Barthelmy et al. 2005). The BAT is sensitive in the 15-150 keV range and, while searching for bursts and monitoring hard X-ray transients, it performs an all-sky hard X-ray survey. The *Swift*-BAT 105 month catalog (Oh et al. 2018) reports the results of this survey, including 1632 hard X-ray sources, ∼60% of which are non-beamed AGN in nearby (z < 0.2) galaxies. The BAT energy range is least biased against CT source detection, making the BAT catalogue the ideal tool to perform a complete survey of CT-AGN.

The Clemson Compton-Thick AGN (CCTAGN)[2] project has been targeting Compton-thick AGN candidates within the BAT catalog, with the objective to find and characterize all obscured AGN in the local ($z < 0.05$) Universe. In order to determine the true CT-AGN fraction, a volume-limited sample is needed, to overcome the bias against detection of the faintest sources. Indeed, almost 90% of CT-AGN in the BAT catalogue have been discovered at $z \leq 0.05$ (within ∼ 200 Mpc). In comparison, 90% of the population of unabsorbed and Compton-thin AGN falls within $z \leq 0.12$ (Ricci et al. 2017), a factor 2.4 (in distance) higher.

The BAT catalogue lists 417 AGN within $z \leq 0.05$ (BAT 100-month catalogue; Segreto et al. in prep.). In order to estimate their obscuring column densities, a soft X-ray follow-up (by i.e. *Swift-XRT*, *Chandra*, *XMM-Newton*, or *Suzaku*) is necessary. Using observations in the $0.3 - 10$ keV range, previous works have classified 63 BAT sources within this volume as candidate CT-AGN (e.g. Ricci et al. 2015; Marchesi et al. 2017a,b, Silver et al. 2021). However, the uncertainties associated to most $N_H$ values are fairly large, due to the lack of high-quality data in the range bridging soft X-rays and BAT data (i.e. ≈ 7−15 keV).

The Nuclear Spectroscopic Telescope Array (hereafter *NuSTAR*; Harrison et al. 2013), observing in the range of $3 - 78$ keV, provides a two orders of magnitude better sensitivity than previous telescopes at energies ≥ 10 keV. This allows one to characterize the properties of the AGN torus (i.e. average $N_H$, inclination angle, covering factor), which mainly affect the reflected emission of the AGN; the so-called 'Compton-hump', at energies ∼ 20 − 40 keV. The addition of *NuSTAR* data allows to break degeneracies between parameters, such as the photon index, the line-of-sight $N_H$, and the reflected emission, thus improving our classification of the sources.

This work is a follow-up on that performed by Marchesi et al. (2018), which presented the analysis of the 38 candidate CT-AGN in the BAT 100 month catalog for which an archival *NuSTAR* observation existed (the largest sample of heavily obscured AGN analyzed with *NuSTAR* so far). The largest study before that contained 11 objects (Masini et al. 2016), and those before focused on single or few sources (e.g. Baloković et al. 2014; Puccetti et al. 2014; Annuar et al. 2015; Bauer et al. 2015; Brightman et al. 2015; Koss et al. 2015; Rivers et al. 2015; Puccetti et al. 2016). This remains true for studies performed since (e.g. Zhao et al. 2019a,b; Turner et al. 2020; Iwasawa et al. 2020). Such studies allow to confirm or rule out the CT nature of a candidate, bringing us closer to deriving the true fraction of CT-AGN in the local Universe.

In this work, we analyse 8 additional CT-AGN candidates selected from the BAT catalogue, for which we were awarded simultaneous *NuSTAR* and *XMM-Newton* data. These are part of the last ten sources in the 63 CT-AGN candidate sample that were still missing *NuSTAR* data, bringing us closer to completing the classification of the full sample. In this work, we decouple the column density in the line of sight, $N_{H,los}$ from the average column density of the torus, $N_{H,av}$, as previous works have shown this strategy provides a better fit to the data (e.g. Marchesi et al. 2019). Furthermore, there is evidence suggesting the AGN torus is a clumpy medium (e.g. Risaliti et al. 2002; Elvis et al. 2004; Markowitz et al. 2014), in which these two values are not necessarily the same. Given this fact, we clarify that we refer to a Compton-thick AGN as one that is Compton-thick in the line of sight.

This work is organized as follows: In Sect. 2 we describe the sample selection and data reduction. In Sect. 3 we describe the X-ray analysis and the models used. In Sect. 4 we present our results and comment on the properties of our sources. Finally, in Sects. 5 and 6, we present our discussion and conclusions, respectively.

## 2. SAMPLE SELECTION AND DATA REDUCTION

The sample in this work is selected from CT-AGN candidates in the Palermo *Swift*-BAT 100-month Catalog (Segreto et al. in prep), detected in the local (z<0.05,

---

[1] At $z > 1$ the 'Compton-hump' is redshifted into the lower-energy range (< 10 KeV), which is sampled by various soft X-ray observatories (e.g. Buchner et al. 2015; Lanzuisi et al. 2015)

[2] https://science.clemson.edu/ctagn/

D≲200 Mpc) Universe. All sources in the sample have been previously analyzed using a combination of *Swift*-BAT and 2−10 keV data, with the results of best-fit models for their X-ray emission classifying them as CT-AGN candidates. However, their column density determination is compatible with $N_{\rm H} < 10^{24}$ cm$^{-2}$ within errors.

The classification of most sources as CT-AGN candidates was reported in Ricci et al. (2015), who fitted the sources using the physical torus model of Brightman & Nandra (2011). ESO 112−G006 is the only source in the sample that was instead analyzed by our group and the results of the preliminary analysis are unreported. In order to confirm (or rule out) their CT nature, we were awarded simultaneous *NuSTAR* and *XMM-Newton* observation time in Cycles 18 and 5 respectively, as part of a *NuSTAR* Large Program (NuSTAR proposal ID: 5197; PI: Marchesi). The details of these observations can be found in Table 1: overall, we were granted ∼500 ks of observations with *NuSTAR* and ∼200 ks with *XMM-Newton* (pre-data cleaning). We were also granted time to observe two other sources (NGC 3081 and ESO 565-G019), selected in the same way. These objects also have archival *Chandra* data and are analyzed in a companion paper (Traina et al. in preparation).

Table 1 also lists previous *XMM-Newton* observations taken from the archive for NGC 6552 and MRK 662, which we used to constrain variability either in the flux or the column density of the sources. Both sources had one additional observation, which we did not use due to high percentage of flaring time, which resulted in poor statistics. All sources have additional *Swift*-XRT observations which, due to low count statistics, do not allow to constrain any possible $N_{\rm H}$ variability.

The data retrieved for both *NuSTAR* Focal Plane Modules (FPMA and FPMB; Harrison et al. 2013) were processed using the NuSTAR Data Analysis Software (NUSTARDAS) v1.8.0. The event data files were calibrated running the nupipeline task using the response file from the Calibration Database (CALDB) v. 20200612. With the nuproducts script, we generated both the source and background spectra, and the ancillary and response matrix files. For both focal planes, we used a circular source extraction region with a 75″ diameter (corresponding to ∼ 80% encircled energy fraction) centered on the target source (except for NGC 6552, for which a 50″ region was used due to high background counts). For the background, we used an annular extraction region (inner radius 100″, outer radius 160″) surrounding the source, excluding any resolved sources. The *NuSTAR* spectra have then been grouped with at least 20 counts per bin.

We reduced the *XMM-Newton* data using the SAS v18.0.0, cleaning for flaring periods and adopting standard procedures. The source spectra were extracted from a 30″ circular region (corresponding to ∼ 85% encircled energy fraction for EPIC-PN), while the background spectra were obtained from a circle that has a radius 45″ located near the source and is not contaminated by nearby objects. Each spectrum has been binned with at least 15 counts per bin.

We fitted our spectra using the XSPEC software (Arnaud 1996, in HEASOFT version 6.26.1), taking into account the Galactic absorption measured by Kalberla et al. (2005). We used Anders & Grevesse (1989) cosmic abundances, fixed to the solar value, and the Verner et al. (1996) photoelectric absorption cross-section. The luminosity distances are computed assuming a cosmology with $H_0$=70 km s$^{-1}$ Mpc$^{-1}$, and $\Omega_\Lambda$=0.73. We used $\chi^2$ as the fitting statistic.

## 3. X-RAY SPECTRAL ANALYSIS

In this section, we describe the different torus models used to fit the X-ray data of each galaxy. Results of the X-ray spectral analysis of each source can be found in Sect. 4. All sources have been fit in the range from 0.6 keV to 25−55 keV, with the higher energy limit depending on the point in which *NuSTAR* data is overtaken by the background. For every source, all models have been consistently applied to the same energy range.

We add a thermal emission component, mekal (Mewe et al. 1985; Kaastra 1992; Liedahl et al. 1995), to all torus models, which is necessary to account for the soft excess below ∼1 keV. We report the best-fit parameters of this model in Sect. 4, but we note that the gaseous material surrounding the AGN and within the galaxy is likely to be multi-phase and complex (see e.g. Torres-Albà et al. 2018), and therefore the derived temperature values, $kT$, should not be taken as accurate estimates of the physical properties of the galaxy.

We also add an additional scattered component, to characterize the intrinsic powerlaw emission of the AGN that either leaks through the torus without interacting with it, or interacts with the material via elastic collisions. This component is set equal to the intrinsic powerlaw, multiplied by a constant, $F_{\rm s}$, that represent the fraction of scattered emission (typically of the order of few percent, or less).

Finally, we take into account the Galactic absorption via the inclusion of a photoelectric absorption (phabs) component to the models.

### 3.1. *MYTorus in Coupled Configuration*

The MYTorus model of Murphy & Yaqoob (2009) assumes an isotropic X-ray emission within a uniform, neu-

Table 1. Source observation details

| Swift-BAT ID | Source Name | R.A. [deg (J2000)] | Decl. [deg (J2000)] | z | Telescope | Obs ID | Exp. Time [ks] | Obs Date |
|---|---|---|---|---|---|---|---|---|
| (1) | (2) | (3) | (4) | (5) | (6) | (7) | (8) | (9) |
| J0030.9−5901 | ESO 112-G006 | 00 30 43.83 | -59 00 25.87 | 0.02885 | *NuSTAR* | 60561038002 | 56.0 | 2019 Nov 6 |
| | | | | | *XMM-Newton* | 0852180101 | 19.1 | 2019 Nov 6 |
| J0105.4-4211 | MCG-07-03-007 | 01 05 26.81 | -42 12 58.3 | 0.02988 | *NuSTAR* | 60561039002 | 54.7 | 2019 Nov 28 |
| | | | | | *XMM-Newton* | 0852180201 | 18.4 | 2019 Nov 28 |
| J0623.7−3213 | ESO 426-G002 | 06 23 46.42 | -32 12 59.51 | 0.02243 | *NuSTAR* | 60561040002 | 52.5 | 2019 Oct 9 |
| | | | | | *XMM-Newton* | 0852180301 | 21.1 | 2019 Oct 9 |
| J0656.2-4919 | LEDA 478026 | 06 56 11.95 | -49 19 50.0 | 0.04100 | *NuSTAR* | 60561041002 | 55.5 | 2020 Feb 23 |
| | | | | | *XMM-Newton* | 0852180401 | 21.2 | 2020 Feb 23 |
| J0807.9+3859[a] | MRK 622 | 08 07 40.99 | +39 00 15.26 | 0.02335 | *NuSTAR* | 60561042002 | 54.2 | 2019 Sep 28 |
| | | | | | *XMM-Newton* | 0138951401 | 6.9 | 2003 May 5 |
| | | | | | *XMM-Newton* | 0852180501 | 8.4 | 2019 Sep 28 |
| J1800.0+6636 | NGC 6552 | 18 00 07.25 | +66 36 54.35 | 0.02656 | *NuSTAR* | 60561046002 | 48.6 | 2019 Aug 20 |
| | | | | | *XMM-Newton* | 0112310801 | 7.4 | 2002 Oct 18 |
| | | | | | *XMM-Newton* | 0852180901 | 11.0 | 2019 Aug 20 |
| J1253.3-4138 | ESO 323-G032 | 12 53 20.31 | -41 38 08.3 | 0.01600 | *NuSTAR* | 60561045004 | 50.2 | 2020 Feb 2 |
| | | | | | *XMM-Newton* | 0852181201 | 18.3 | 2020 Feb 2 |
| J2307.8+2242 | CGCG 475-040 | 23 07 48.86 | +22 42 37.0 | 0.03473 | *NuSTAR* | 60561047002 | 55.8 | 2019 Nov 29 |
| | | | | | *XMM-Newton* | 0852181001 | 21.7 | 2019 Nov 30 |

**Notes:** (1): ID from the Palermo BAT 100 months Catalog (Marchesi et al. 2019). (2): Source name. (3) and (4): R.A. and decl. (J2000 Epoch). (5): Redshift. (6): Telescope used in the analysis. (7): Observation ID. (8): Exposure time, in ks. *XMM-Newton* values are reported for EPIC-PN, after cleaning for flares. (9): Observation date. a) ID from the 105-month catalogue of Oh et al. (2018), as the source is not detected in the Palermo BAT catalogue.

tral (cold) torus. The half-opening angle of the torus is fixed to 60°.

The MYTorus model is decomposed into three different components: an absorbed line-of sight emission, a reflected continuum, and a fluorescent line emission. These components are linked to each other with the same normalization, absorbing column density (the model yields the equatorial column density of the torus, $N_{\rm H,eq}$) and inclination angle $\theta_{\rm i}$. The inclination angle is measured from the axis of the torus, so that $\theta_{\rm i}$=0° represents a face-on AGN, and $\theta_{\rm i}$=90° an edge-on one.

The line of sight column density of the torus can be obtained from the equatorial value as

$$N_{\rm H,los} = N_{\rm H,eq} \times (1 - 4 \times cos(\theta_{\rm i})^2)^{1/2} \quad (1)$$

and the average torus column density is determined by the given configuration as $N_{\rm H,av} = \pi/4 \times N_{H,eq}$, and cannot be decoupled (i.e. fit separately) from the line of sight value.

Both the reflected continuum and line emission can be weighted via two additional constants, $A_{\rm S}$ and $A_{\rm L}$, respectively. When left free to vary, these can account for differences in the actual geometry (compared to the specific model assumptions used in the original calculations) and time delays between direct, scattered and fluorescent line photons.

In XSPEC this model configuration is as follows,

$$Model = C * phabs *$$
$$(mekal + mytorus\_Ezero\_v00.fits * zpowerlw +$$
$$A_{\rm S} * mytorus\_scatteredH500\_v00.fits +$$
$$A_{\rm L} * mytl\_V000010nEp000H500\_v00.fits$$
$$+F_{\rm s} * zpowerlw) \quad (2)$$

where C is a cross-calibration constant between different instruments, or a cross-normalization constant between different observations.

### 3.2. MYTorus in Decoupled Configuration

The MYTorus model can also be used in 'decoupled configuration' (Yaqoob 2012), so as to better represent the emission from a clumpy torus. While the model has the exact same assumptions, a better description of the data is possible when decoupling the line-of-sight emission from the reflection component. That is, $N_{H,los}$ and $N_{H,av}$ are not fixed to the same value, allowing the flexibility of having a particularly dense line of sight in a (still uniform) Compton-thin torus, or vice versa.

In this configuration, the line of sight inclination angle is always fixed to $\theta_i$=90º, and two reflection and line components are included, one with $\theta_i$=90º (forward scattering) and weighted with $A_{S,L90}$ and one with $\theta_i$=0º (backward scattering) and weighted with $A_{S,L0}$. Note that in this configuration $\theta_i$ is no longer a variable, although the ratio between $A_{S,L0}/A_{S,L90}$ can give a qualitative idea of which direction reflection predominantly comes from, and therefore of the relative orientation of the AGN[3].

In XSPEC this model configuration is as follows,

$$Model = C * phabs *$$
$$(mekal + mytorus\_Ezero\_v00.fits * zpowerlw +$$
$$A_{S,0} * mytorus\_scatteredH500\_v00.fits +$$
$$A_{L,0} * mytl\_V000010nEp000H500\_v00.fits +$$
$$A_{S,90} * mytorus\_scatteredH500\_v00.fits +$$
$$A_{L,90} * mytl\_V000010nEp000H500\_v00.fits +$$
$$+ F_s * zpowerlw) \quad (3)$$

We fix $A_{S,90} = A_{L,90}$ and $A_{S,0} = A_{L,0}$. In the default MYTorus decoupled scenario, we freeze all these constants to unity. However, we also include the results of thawing them both, which in some cases improves fit quality. We call this third model 'MYTorus decoupled free'.

### 3.3. BORUS02

The other model used in this work is borus02 (Baloković et al. 2018). This model also assumes a uniform torus, but the opening angle is not fixed, and different geometries can be considered through the covering factor, $C_F$ parameter ($C_f \in [0.1, 1]$). The model only includes a reflection component, which contains both the continuum and lines, meaning that an absorbed line-of-sight component must be added.

By default, we use this model in a decoupled configuration, with $N_{H,los}$ and $N_{H,av}$ set to vary independently.

---

[3] We note that the normalization of both 0 and 90-degree scattering components is linked to that of the intrinsic continuum, and therefore it is necessary to leave both $A_{S,0}$ and $A_{S,90}$ free to vary3.

An advantage of this model, aside from including a variable covering factor, is that $\theta_i$ (with $\theta_i \in [18º-87º]$) can still be fitted in a decoupled configuration. borus02 also includes a high-energy cutoff and iron abundance as free parameters, although we freeze them to 500 keV (consistent with the default setting in MYTorus) and 1, respectively, due to our inability to constrain them. We note that some works estimate lower ($\sim$ 200 keV) high-energy cutoffs (e.g., Ricci et al. 2017; Ananna et al. 2020). However, the most recent work of Baloković et al. (2020) focuses on the local obscured AGN population (i.e. more similar to our sample properties) and places the average cutoff at $\sim$ 300 keV. They also show that systematic uncertainties allow for a relatively wide range, which marginally includes 500 keV. Note, also, that with NuSTAR data reaching up to $25 - 55$ keV energies, the different high-energy cutoff values do not impact our results. CGCG 475-040 is the exception to this rule, as it required a very low high-energy cutoff to adequately fit the data.

In XSPEC this model configuration is as follows,

$$Model = C * phabs * (mekal+$$
$$borus02\_v170323a.fits + zphabs * cabs * zpowerlw$$
$$+ F_s * zpowerlaw), \quad (4)$$

where zphabs and cabs are the photoelectric absorption and Compton scattering, respectively, applied to the line-of-sight component.

## 4. RESULTS

In this section we describe the results obtained via X-ray spectral fitting, using simultaneous XMM-Newton and NuSTAR data, and compare them to previous determinations (when available). We note that the 'MYTorus coupled' fits to the data are generally worse, statistically speaking, and often in disagreement with the results of other models (with the two exceptions justified in the text). This is a result of not allowing the $N_{H,los}$ and $N_{H,av}$ to vary independently, which may yield an averaged-out value. Due to this fact, the discussion of the fitting results for each source does not take the 'MYTorus coupled' model into consideration. We will further discuss the validity of this model in Sect. 5.

An example table showing the best-fit parameters for ESO 112-G006, Table 2, is presented in the text. Tables for the rest of sources can be found in Appendix A. The tables also give our estimation of flux (observed) and luminosity (intrinsic) derived using each best-fit model and the cflux and clumin XSPEC convolution components. The equivalent width of the iron $K_\alpha$ line (EW) is computed as described in Marchesi et al. (2018). All errors reported are at a 90% confidence level unless otherwise stated.

Likewise, we show plots of the MYTorus decoupled free and borus02 fits to the data for ESO 112−G006 in Fig. 1, and the rest of figures can be found in Appendix B.

### 4.1. *ESO 112-G006*

This source was first reported as a CT-AGN candidate by Ricci et al. (2015), who obtained a value of log$N_H$ = 24.03$^{+0.40}_{-0.24}$ based on *XMM-Newton* and *Swift*-BAT observations. There are no optical classifications on its activity type, but its optical spectrum (Jones et al. 2009) does not present any broad emission lines. Results of the fitting can be found in Table 2.

This source is well-fitted by all models except for MYTorus decoupled. Note that, generally, MYTorus decoupled provides a better fit than the coupled version. This is likely because, for this particular source, there is no contribution coming from face-on reflection, and the mentioned model assumes $A_{S0} = A_{S90} = 1$. Indeed, MYTorus decoupled free is a better fit, showing a clear predominance of forward reflection, which agrees with the borus02 inclination angle being large ($\cos(\theta_i) < 0.3$).

All models agree that this source is observed through a Compton-thin line of sight ($N_{H,los} = 0.47 - 0.67 \times 10^{24}$ cm$^{-2}$), while the average torus material is denser, and even CT ($N_{H,av} = 0.71 - 2.42 \times 10^{24}$ cm$^{-2}$). According to the borus02 best-fit model, this CT torus would be geometrically thin, with a relatively small covering factor ($C_F = 0.21^{+0.32}_{-0.09}$). The photon index lies in the range $\Gamma = 1.40 - 1.83$, when considering all the different models.

### 4.2. *MCG-07-03-007*

MCG-07-03-007[4] was first reported as a CT-AGN candidate by Ricci et al. (2015), who obtained a value of log$N_H$ = 24.18$^{+0.12}_{-0.35}$ based on *Swift*-XRT and *Swift*-BAT observations. The source is optically classified as a Sy2 (Baumgartner et al. 2013). Results of the fitting can be found in Table 5.

All models are in good agreement for the description of the source. It is marginally Compton-thin in the line of sight ($N_{H,los} = 0.73 - 0.97 \times 10^{24}$ cm$^{-2}$), with $\Gamma \sim 1.8$ and a Compton-thick torus. The results of 'MYTorus decoupled free' have larger uncertainties, likely due to the fact that the addition of two more parameters is not required by the fit. borus02 gives a best fit with a covering factor of 0.6, just barely intercepted by the line of sight.

### 4.3. *ESO 426-G002*

This source was selected as a candidate CT-AGN based on our own *Swift*-XRT and *Swift*-BAT analysis, which is unreported in any previous publications. Optically, it is classified as a Sy2 (Baumgartner et al. 2013). Results of the fitting can be found in Table 6.

This source is clearly best-fit by MYTorus decoupled free and borus02, with remarkably similar results, which (except for the photon index) do not differ significantly from the MYTorus decoupled best-fit model. According to our analysis, this source is borderline CT in the line of sight, and CT in the average torus material ($N_{H,los} = 0.92 - 1.09 \times 10^{24}$ cm$^{-2}$, $N_{H,av} = 2.86 - 4.64 \times 10^{24}$ cm$^{-2}$). borus02 can constrain the covering factor and inclination angle with high accuracy thanks to the clear dominance of the reflection component. For this source, the dominance of forward reflection (according to MYTorus decoupled free) would lead us to believe that the source has a large inclination angle, but borus02 results place it at $\theta_i \sim 30°$. It could be that, given the large covering factor ($C_F = 0.97^{+0.02}_{-0.03}$), most of the reflection comes through the torus, which MYTorus interprets as 90° reflection, regardless of the actual direction.

### 4.4. *LEDA 478026*

This source was first reported as a CT-AGN candidate by Ricci et al. (2015), who obtained a value of log$N_H$ = 24.03$^{+0.30}_{-0.10}$ based on *Swift*-XRT and *Swift*-BAT observations. Optically, it is classified as a Sy2 (Baumgartner et al. 2013). Results of the fitting can be found in Table 7.

For this source, MYTorus decoupled free and borus02 are in strong agreement, fitting it as Compton-thick in the line of sight ($N_{H,los} \sim 1.45 \times 10^{24}$ cm$^{-2}$), with a lower average torus column density ($N_{H,av} \sim 0.33 \times 10^{24}$ cm$^{-2}$). With an estimated covering factor of only $C_f = 0.15$, this source is likely to have a very patchy torus. In this scenario, the low covering factor should not be interpreted geometrically (i.e., like in a thin "disk-like" torus) but rather physically, meaning that the surrounding clouds obscure only a small fraction of the available volume.

Contrary to this, MYTorus decoupled gives a different estimation for all the significant parameters, with a harder photon index, Compton-thin line of sight and Compton-thick torus. However, the fit is statistically

---

[4] We note that this source can be easily confused with UGC 0058, as it is mistakenly named MCG 07-03-007 or MGC 07-03-007 on occasion (without the minus sign in front) which in SIMBAD or NED redirects to the mentioned source. The correct position and redshift of the analysed source can be found in Table 1.

Table 2. X-ray fitting results of ESO 112-G006

| Model | MYTorus (Coupled) | MYTorus (Decoupled) | MYTorus (Decoupled free) | borus |
|---|---|---|---|---|
| red $\chi^2$ | 1.08 | 1.27 | 1.08 | 1.02 |
| $\chi^2$/d.o.f. | 276.61/256 | 325.42/256 | 274.29/254 | 260.37/254 |
| $kT$ | $0.88^{+0.17}_{-0.11}$ | $0.87^{+0.16}_{-0.10}$ | $0.88^{+0.17}_{-0.11}$ | $0.86^{+0.17}_{-0.12}$ |
| $\Gamma$ | $1.46^{+0.07}_{-}$ | $1.58^{+0.04}_{-}$ | $1.48^{+0.08}_{-}$ | $1.60^{+0.23}_{-0.13}$ |
| $N_{H,los}$ | $0.57^{+0.06}_{-0.07}$ | $0.53^{+0.03}_{-0.06}$ | $0.57^{+0.05}_{-0.05}$ | $0.63^{+0.04}_{-0.06}$ |
| $N_{H,eq}$ | $0.57^{+0.06}_{-0.03}$ | – | – | – |
| $N_{H,av}$ | – | $1.99^{+0.30}_{-0.86}$ | $1.99^{+0.43}_{-0.60}$ | $1.11^{+0.74}_{-0.40}$ |
| $A_{S90}$ | – | 1* | $0.68^{+0.59}_{-0.35}$ | – |
| $A_{S0}$ | – | 1* | 0* | – |
| $C_F$ | – | – | – | $0.21^{+0.32}_{-0.09}$ |
| $\cos(\theta_i)$ | $0.05^{+0.13}_{-}$ | – | – | $0.15^{+0.15}_{-}$ |
| $F_s$ ($10^{-3}$) | $0.81^{+1.54}_{-}$ | 0* | $0.74^{+1.41}_{-}$ | $0.88^{+1.09}_{-}$ |
| Norm ($10^{-4}$) | $6.16^{+1.65}_{-1.01}$ | $6.61^{+0.79}_{-2.86}$ | $6.93^{+1.75}_{-1.62}$ | $10.1^{+8.5}_{-3.2}$ |
| EW [keV] | $0.11^{+0.05}_{-0.05}$ | $0.14^{+0.05}_{-0.05}$ | $0.11^{+0.05}_{-0.05}$ | – |
| Flux (2−10 keV) [$10^{-13}$] | $4.87^{+0.31}_{-0.31}$ | $4.50^{+0.29}_{-0.29}$ | $4.90^{+0.33}_{-0.33}$ | $4.92^{+0.31}_{-0.31}$ |
| Flux (10−40 keV) [$10^{-12}$] | $4.98^{+0.15}_{-0.15}$ | $5.15^{+0.15}_{-0.15}$ | $4.97^{+0.15}_{-0.15}$ | $4.74^{+0.14}_{-0.14}$ |
| $L_{intr}$ (2-10 keV) [$10^{42}$] | $5.59^{+0.51}_{-0.51}$ | $4.42^{+0.46}_{-0.46}$ | $6.23^{+0.52}_{-0.52}$ | $8.01^{+0.64}_{-0.64}$ |
| $L_{intr}$ (15-55 keV) [$10^{43}$] | $1.52^{+0.53}_{-0.54}$ | $1.06^{+0.65}_{-0.65}$ | $1.60^{+0.53}_{-0.53}$ | $1.41^{+0.46}_{-0.46}$ |
| counts | 7053 | | | |

**Notes:**
red $\chi^2$: reduced $\chi^2$
$\chi^2$/d.o.f.: $\chi^2$ over degrees of freedom.
$kT$: mekal model temperature, in units of keV.
$\Gamma$: Powerlaw photon index.
$N_{H,los}$: Line-of-sight torus hydrogen column density, in units of $10^{24}$ cm$^{-2}$.
$N_{H,av}$: Equatorial torus hydrogen column density, in units of $10^{24}$ cm$^{-2}$.
$N_{H,av}$: Average torus hydrogen column density, in units of $10^{24}$ cm$^{-2}$.
$A_{S90}$: Constant associated to the reflection component, edge-on.
$A_{S0}$: Constant associated to the reflection component, face-on.
$C_F$: Covering factor of the torus, $\in[0.1-1]$.
$\cos(\theta_i)$: cosine of the inclination angle. $\cos(\theta_i)=1$ represents a face-on scenario.
$F_s$: Fraction of scattered continuum
Norm: Normalization of the AGN emission.
EW: Equivalent width of the neutral iron K-alpha line.
Fluxes (observed) are given in units of erg s$^{-1}$ cm$^{-2}$.
Luminosities (intrinsic) are given in units of erg s$^{-1}$.
Total net counts used for fitting: *XMM-Newton* in the 0.6−9 keV band, and *NuSTAR* from 3 to 25-55 keV (depending on the source. See full range for each source in the plots shown in the Appendix)
*: Variable fixed to the respective value.
Unreported upper/lower limits for any variable represent the inability of the model to provide them (i.e. parameter is compatible within 90% error with the model hard limits) worse, and the `MYTorus` decoupled free results point toward a strong dominance of only forward reflection (agreeing with the `borus02` edge-on viewing angle). This leads us to favor the former scenario, given how `MYTorus` decoupled is limited by forcing $A_{S,L90} = A_{S,L0} = 1$.

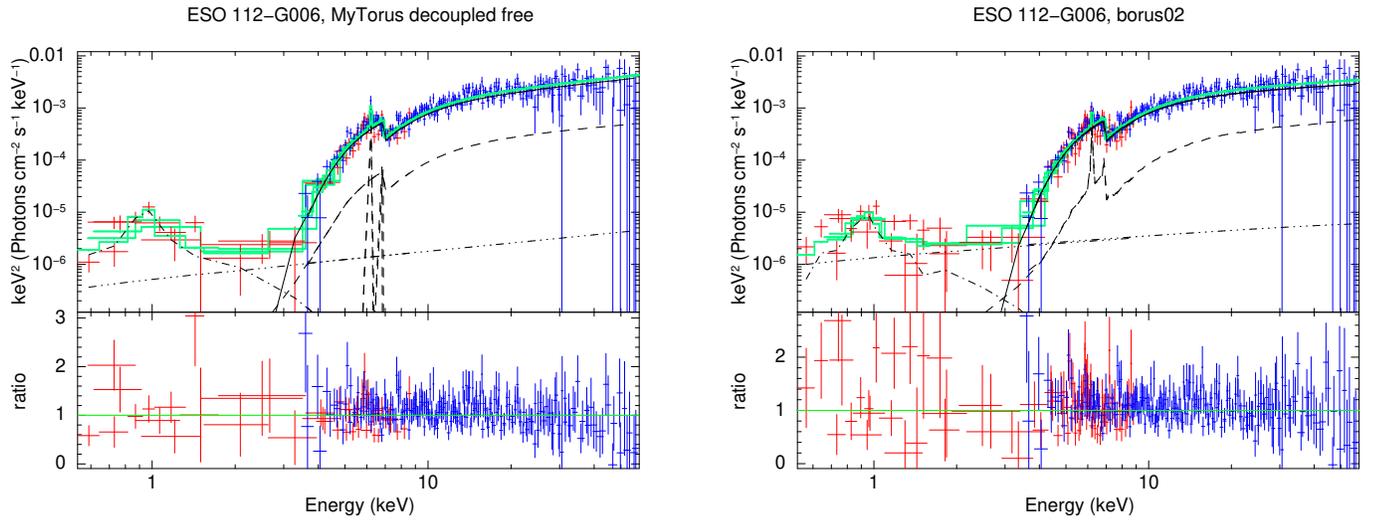

**Figure 1.** X-ray spectral fitting (unfolded) of ESO 112−G006 using MYTorus in 'decoupled free' configuration (left) and borus02 (right). In both plots, *XMM-Newton* and *NuSTAR* data are plotted in red and blue crosses, respectively. The best-fit convolution model is shown in a solid, green line. The different components are shown in black lines: line-of-sight emission (solid), reflected emission (dashed in borus02. For MYTorus, the 90° reflection is shown in a dashed line, and the 0° reflection in a dotted line), scattered emission (dash-dot-dot-dot) and soft thermal emission (dash-dot). Note that this particular source does not show any 0° reflection in the MYTorus model.

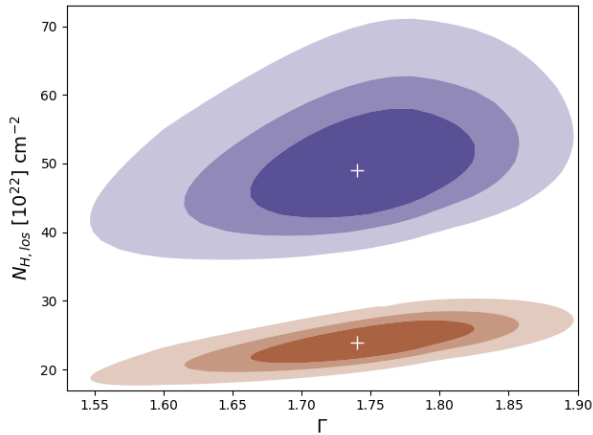

**Figure 2.** Confidence contours at 68, 90, 95% levels, of photon index and line-of-sight hydrogen column density determinations, for the two *XMM-Newton* observations of MRK 622. In blue, May 5th 2003, and in brown, 28th September 2019.

### 4.5. *MRK 622*

This source was first reported as a CT-AGN candidate by Ricci et al. (2015), who obtained a value of $\log N_\mathrm{H} = 24.29^{-}_{-0.30}$[5] based on *XMM-Newton* and *Swift*-BAT observations. Optically, it is classified as a Sy2 (Véron-Cetty & Véron 2006). Results of the fitting can be found in Tables 8 and 9, the latter of which includes the second *XMM-Newton* observation, taken from the archive.

In the case of MRK 622, the `MYTorus` model does not show a statistical improvement when adding additional free parameters. All models place this source as having a CT torus ($N_\mathrm{H,av} = 0.63 - 3.11 \times 10^{24}$ cm$^{-2}$), while having a much lower line of sight column density ($N_\mathrm{H,los} = 0.15 - 0.29 \times 10^{24}$ cm$^{-2}$). There is a disagreement between the best-fit photon index value between `MYTorus` models and `borus02` model, with the first set at $\Gamma = 1.54^{+0.14}_{-}$, and the second at $\Gamma = 1.74^{+0.17}_{-0.19}$. The determination of torus properties, such as covering factor and opening angle, is made difficult by the fact that the reflection component is subdominant with respect to the line of sight (hence the unconstrained values). This is also likely to be the reason for `MYTorus` decoupled free to not show a statistical improvement of fit, as the added parameters model reflected emission.

When adding the archived data, we introduce a cross-normalization constant, $C$, and a different line-of-sight hydrogen column density, $N_\mathrm{H,los2}$, and leave both free to vary. The addition does not result in significant changes or incremented agreement between the different models. However, leaving $N_\mathrm{H,los2}$ free to vary results in a best-fit value of $N_\mathrm{H,los2} = 0.39 - 0.69 \times 10^{24}$ cm$^{-2}$, which is incompatible within the errors with the best-fit value for $N_\mathrm{H,los}$, for all models (see Fig. 2). Therefore, we conclude that this source presents line-of-sight $N_\mathrm{H}$ variability at different epochs.

### 4.6. *NGC 6552*

This source was first reported as a CT-AGN candidate by Ricci et al. (2015), who obtained a value of $\log N_\mathrm{H} = 24.05^{+0.35}_{-0.22}$ based on *XMM-Newton* and *Swift*-BAT observations. Optically, it is classified as a Sy2 (Lin et al. 2012). Results of the fitting can be found in Tables 10 and 11, the latter of which includes the second *XMM-Newton* observation, taken from the archive.

All models agree that this source has a CT line-of-sight ($N_\mathrm{H,los} = 1.42 - 3.16 \times 10^{24}$ cm$^{-2}$) within a Compton-thin torus ($N_\mathrm{H,av} = 0.30 - 0.55 \times 10^{24}$ cm$^{-2}$). The range of photon index values is relatively large ($\Gamma = 1.53 - 2.11$), although both `MYTorus` decoupled and decoupled-free are compatible with the `borus02` results within errors. The ratio between $A_{S,90}$ and $A_{S,0}$ suggests a predominance of forward reflection, which is compatible with the observation angle derived by `borus02`, $\theta_\mathrm{i} \sim 75°$.

When adding the archived observation, the data quality did not allow to constrain $N_\mathrm{H,los2}$, and its value was compatible with that of $N_\mathrm{H,los}$. Therefore, the results presented in Table 11 have them fixed to be the same value. Both this and the fact that $C$ is compatible with 1 makes us conclude this source does not present variability between the two analysed observations.

The addition of this second set of *XMM-Newton* data improves the overall agreement between the three models with good fitting statistics, and in particular between `MYTorus` decoupled free and `borus02`. The qualitative description of the results would remain the same, with reduced uncertainty; $N_\mathrm{H,los} = 1.42 - 2.56 \times 10^{24}$ cm$^{-2}$, $N_\mathrm{H,av} = 0.34 - 0.63 \times 10^{24}$ cm$^{-2}$, $\Gamma = 1.57 - 1.84$.

### 4.7. *ESO 323-G023*

This source was first reported as a CT-AGN candidate by Ricci et al. (2015), who obtained a value of

---

[5] No upper error available. Likewise, Unreported upper/lower limits for any variable represent the inability of the used model to provide them (i.e. parameter is compatible within 90% error with the model hard limits)

log$N_\mathrm{H}$ = $24.79^{-}_{-0.40}$ based on *Swift*-XRT and *Swift*-BAT observations. It is optically classified as Sy2 in Bird et al. (2007). This source has an additional *XMM-Newton* +*NuSTAR* joint observation (PI:Marchesi, Obsid: 0852180801, 60561045002) that we do not use due to an error in the datataking. Results of the fitting can be found in Table 12.

All models fit this source with a slightly soft photon index ($\Gamma \sim 2.0$) and a Compton-thick line of sight, $N_{\mathrm{H,los}} = 1.12 - 3.21 \times 10^{24}$ cm$^{-2}$ (ignoring the MyTorus unconstrained result, as the statistics do not justify the addition of two extra free parameters). Also in agreement, they place the average torus column density to be slightly lower, $N_{\mathrm{H,av}} = 0.49 - 1.79 \times 10^{24}$ cm$^{-2}$. Again, borus02 gives a best fit with a covering factor of 0.6, just barely intercepted by the line of sight.

### 4.8. *CGCG 475-040*

This source was first reported as a CT-AGN candidate by Ricci et al. (2015), who obtained a value of log$N_\mathrm{H} = 24.20^{+0.30}_{-0.20}$ based on *Swift*-XRT and *Swift*-BAT observations. It is optically classified as Sy2 in Parisi et al. (2014). Results of the fitting can be found in Table 13.

This source is best-fit with a rather soft photon index, $\Gamma = 1.9 - 2.60$, and a large value of the average hydrogen column density when using the MYTorus model. The borus02 best-fit of the *NuSTAR* +*XMM-Newton* data has two possible configurations fitting the data equally well: 1) a soft photon index, with a dense torus that has a large ($\sim$90%) covering factor, and is viewed at a small inclination angle; 2) a photon index frozen to 1.8, with a very patchy torus (low average column density and covering factor, yet CT in the line of sight). As both options have the same reduced $\chi^2$ we cannot say which model is superior.

To try and disentangle this degeneracy, we added *Swift*-BAT data to the spectrum[6], as shown in Fig. 13. We used borus02 to fit the data, as it is the only model providing an estimate for the torus covering factor, which we are interested in constraining within the two possible options mentioned above. With the addition of the BAT data, the best-fit model favors a scenario with a very dense torus of large covering factor, through which we observe the AGN through an underdense region (since $N_{H,los} = 1.60^{+0.23}_{-0.15} \times 10^{24}$ cm$^{-2}$). Although the average torus density is capped at the maximum possible value (log$N_{H,av} = 25.5$), we note that it is compatible with being only a factor $\sim 1.5$ larger than that of the line of sight.

Interestingly, the BAT data can be adequately fit only when considering a cross-normalization factor between the *NuSTAR* +*XMM-Newton* data and the BAT data ($C = 2.38^{+0.49}_{-0.48}$). This implies our joint observation took place in a low-flux state of the source. It is also necessary to leave the high-energy cutoff free to vary, for which we obtain $E_{\mathrm{cut}} = 21.0^{+17.7}_{-}$ keV. This value, while being low, is not unprecedented (see cutoff energy distribution of Swift-BAT sources; Ricci et al. 2017; Ananna et al. 2019). However, we caution that such a low high-energy cutoff can be spurious when $N_\mathrm{H}$ is high and data quality is not exceptional (see discussion in e.g. Baloković et al. 2020).

All models classify the source as CT, with $N_{H,los} > 10^{24}$ cm$^{-2}$. MYTorus decoupled free, despite having a CT best-fit value, is also compatible with a Compton-thin scenario within errors. However, we note that this model is statistically worse than MYTorus decoupled despite having two additional free parameters. This likely means MYTorus decoupled free has too many free parameters, which increases degeneracy and results in less reliable results.

## 5. DISCUSSION

We classify a source as CT-AGN when its best-fit value for the line-of-sight hydrogen column density is $N_{\mathrm{H,los}} \geq 10^{24}$ cm$^{-2}$. This corresponds to five out of the eight sources analyzed in this work (NGC 6552, ESO 426-G002, CGCG 475-040, ESO 323-G023, LEDA 478026). Note that one of them, ESO 426-G002 is still compatible with having $N_\mathrm{H}$ slightly below this threshold at 90% uncertainty. The other sources, although Compton-thin, are still heavily obscured. Table 3 summarizes the best-fit borus02 parameters for the sample analyzed in this work.

### 5.1. *Compton-thick sources in the local Universe*

Ricci et al. (2015) provided a list of CT-AGN candidates in the 70-month BAT catalogue, based on joint BAT and soft X-rays analysis (the best available data out of *XMM-Newton*, *Chandra*, *Suzaku*, *Swift-XRT*, and *ASCA*). Out of a total of 55, 50 fall within $z < 0.05$. Based on the Palermo 100 catalog (Cusumano et al. 2014, Segreto et al. in prep.), other works (Marchesi et al. 2017a,b, Silver et al. 2021) have added to the list of possible CT-AGN within the BAT catalogue. This comes to a total of 63 CT-AGN candidates at $z \leq 0.05$.

Marchesi et al. (2018) analysed 38 of these sources using *NuSTAR* data, which is key to disentangling the

---

[6] We note that for no other source the addition of *Swift*-BAT data represented an improvement to the joint *XMM-Newton* +*NuSTAR* fit. Cutoff energy estimations for other sources in this work can be found in Ricci et al. (2017), who could not estimate $E_\mathrm{cut}$ for CGCG 475-040, possibly due to the mentioned parameter degeneracies.

**Table 3.** Best-fit `borus02` parameters for the whole sample

| Source | $\Gamma$ | $N_{\rm H,los}$ | $N_{\rm H,av}$ | $C_{\rm F}$ | $\cos\theta_{\rm i}$ |
|---|---|---|---|---|---|
| | | $[10^{24}\ {\rm cm}^{-2}]$ | $[10^{24}\ {\rm cm}^{-2}]$ | | |
| ESO 112−G006 | $1.60^{+0.23}_{-0.13}$ | $0.63^{+0.04}_{-0.06}$ | $1.11^{+0.74}_{-0.40}$ | $0.21^{+0.32}_{-0.09}$ | $0.15^{+0.15}_{-}$ |
| MCG−07−03−007 | $1.84^{+0.12}_{-0.15}$ | $0.90^{+0.07}_{-0.08}$ | $3.15^{+5.55}_{-0.28}$ | $0.60^{+0.36}_{-0.10}$ | $0.57^{+0.13}_{-0.17}$ |
| ESO 426−G002 | $2.08^{+0.02}_{-0.03}$ | $1.02^{+0.03}_{-0.03}$ | $3.16^{+0.55}_{-0.30}$ | $0.97^{+0.02}_{-0.03}$ | $0.87^{+0.02}_{-0.01}$ |
| LEDA 478026 | $1.72^{+0.07}_{-0.09}$ | $1.44^{+0.16}_{-0.09}$ | $0.34^{+0.11}_{-0.14}$ | $0.15^{+0.05}_{-}$ | $0.05^{+0.23}_{-}$ |
| MRK 622 | $1.74^{+0.12}_{-0.13}$ | $0.24^{+0.03}_{-0.04}$ | $1.50^{+0.65}_{-0.38}$ | $1.00^{-}_{-0.40}$ | $0.84^{-}_{-}$ |
| NGC 6552 | $1.76^{+0.08}_{-0.12}$ | $2.18^{+0.38}_{-0.35}$ | $0.48^{+0.15}_{-0.13}$ | $0.40^{+0.09}_{-0.05}$ | $0.34^{+0.11}_{-0.11}$ |
| ESO 323−G032 | $2.02^{+0.13}_{-0.30}$ | $1.75^{+1.46}_{-0.49}$ | $0.98^{+0.28}_{-0.49}$ | $0.61^{+0.37}_{-0.06}$ | $0.55^{+0.27}_{-}$ |
| CGCG 475−040 | $1.72^{+0.15}_{-0.12}$ | $1.60^{+0.23}_{-0.15}$ | $31.6^{-}_{-29.1}$ | $0.90^{+0.06}_{-0.21}$ | $0.87^{+0.01}_{-0.11}$ |

**Notes:** Parameters are as defined in Table 2. For sources with multiple observations, the best-fit values are taken from fitting them together. For MRK 622, which shows variable $N_{\rm H,los}$, the value listed is that of the joint *NuSTAR* and *XMM-Newton* observation, which has higher count statistics.

degeneracy between the photon index, $\Gamma$, the line-of-sight column density, $N_{\rm H,los}$, and the reflection component (which depends on the average column density, $N_{\rm H,av}$, inclination angle and torus covering factor), and confirmed the CT nature of 17 them (Originally 20 in Marchesi et al. (2018), but the reanalysis performed by Zhao et al. (2020b) detailed in Sect. 5.2 reclassified three of them into Compton-thin). Further works have brought the number of BAT-selected, confirmed CT-AGN at $z \leq 0.05$, to a total of 29 (Koss et al. 2016; Oda et al. 2017; Georgantopoulos & Akylas 2019; Tanimoto et al. 2019; Kammoun et al. 2020; Zhao et al. 2020a,b, Traina et al. in prep.).

In this work, we report 5 additional *NuSTAR*-confirmed CT sources, bringing the current number to 32. This work thus represents a ∼19% increase of confirmed CT-AGN over the previous sample. The full list of *NuSTAR*-confirmed CT-AGN in the BAT catalogue at $z \leq 0.05$, along with the references to their analysis, can be found in our website[7].

According to the latest version of the Palermo BAT catalogue (100-month catalogue, Segreto et al. in prep.), there is a total of 414 BAT-detected AGN within z≤0.05. We note that we are including galaxies lacking an optical classification as AGN (i.e. galaxy, galaxy in pair, galaxy in group, emission-line galaxy and infrared galaxy), given how their bright emission at > 15 keV is difficult to explain through other means. This implies that the number of confirmed CT-AGN within the volume-limited sample is still ∼ 8%, far from model predictions. The most recent estimate, that of Ananna et al. (2019) predicts, based on population synthesis models, a fraction of CT-AGN of ∼ 50%. Note, however, that this prediction is dependent on the flux of the sample considered. Our BAT sample is flux limited and, after applying the pertinent correction, should have a CT-AGN fraction between $27 − 38\%$ according to the model of Ananna et al. (2019)

We also note that our $z < 0.05$ is not necessarily complete. Indeed, we observe a significant trend with redshift of the CT-AGN fraction, which is higher (i.e. closer to predicted values) at lower redshifts. Fig. 3 and Table 4 show the evolution of the observed CT-AGN fraction as a function of redshift, proving that indeed we can recover a ∼ 20% of CT-AGN in the lowest redshift bin. This value lies just below (when taking errors into account) the lower limit of the Ananna et al. (2019) estimations.

These results point to the sample being incomplete even at the lowest possible redshift, likely a result of the BAT flux limit. CT-AGN at higher redshift are likely too faint to be detected by BAT in the first place. We note that the errors listed in Table 4 and shown in Fig. 3 are purely statistical, and do not account for any bias or incompleteness/obscuration corrections. Such predictions are non-trivial and are left for a future study.

In Fig. 4 we show the sources with *NuSTAR* data targeted by our group as part of the Clemson Compton-Thick AGN project, our effort to characterize CT-AGN in the local Universe. Out of the 48 objects analysed by our group, 24 are found to be CT-AGN, which rep-

---
[7] https://science.clemson.edu/ctagn/
We encourage authors to contact us regarding any sources that might be missing.

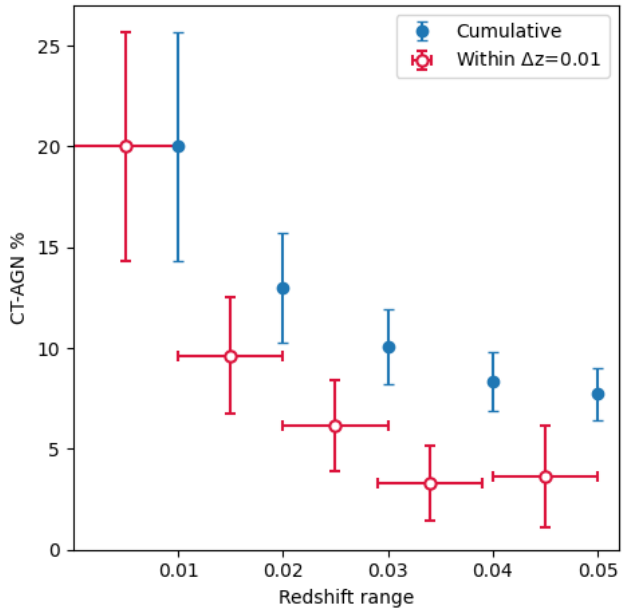

**Figure 3.** CT-AGN fraction within the BAT 100-month catalog (Segreto et al. in prep.) as a function of redshift. The red data represents the fraction within a a given redshift bin of 0.01, while the blue data points correspond to the cumulative value within <z. The computed fractions and total number of sources can be found in Table 4 .

**Table 4.** CT-AGN fraction in the local Universe.

| Redshift | CT-AGN | Total AGN | CT-AGN % |
|---|---|---|---|
| $z \leq 0.01$ | 10 | 50 | $20.0 \pm 5.7$ |
| $z \leq 0.02$ | 20 | 154 | $13.0 \pm 2.7$ |
| $z \leq 0.03$ | 27 | 268 | $10.1 \pm 1.8$ |
| $z \leq 0.04$ | 30 | 359 | $8.4 \pm 1.5$ |
| $z \leq 0.05$ | 32 | 414 | $7.7 \pm 1.3$ |

**Notes:** Observed CT-AGN fraction in the local Universe as a function of redshift. Total AGN include those in the BAT 100-month catalog within a given redshift bin. CT-AGN include those within the mentioned catalog, confirmed by *NuSTAR* as Compton-thick. Errors are binomial.

resents a $\sim 50\%$ of success. This result showcases the need for *NuSTAR* data to confirm CT-AGN candidates, as all sources shown in the figure were compatible with being CT based on soft X-rays and BAT data. It is possible that, without *NuSTAR* data and using simpler, phenomenological models, the selection based on BAT + soft X-ray data cannot properly distinguish between $N_{\rm H,los}$ and $N_{\rm H,av}$, resulting in missclasifications (all Compton-thin sources in this work have a Compton-thick torus). We note that using the phenomenological models on the high-quality data we present here would result in the same effect. However, one can only apply the self-consistent, complex models (that allow to disentangle between $N_{\rm H,los}$ and $N_{\rm H,av}$) in a meaningful way if *NuSTAR* data is available. That is because the reflection component dominates at $\sim 20$ keV, a range in which no other satellite is sensitive enough.

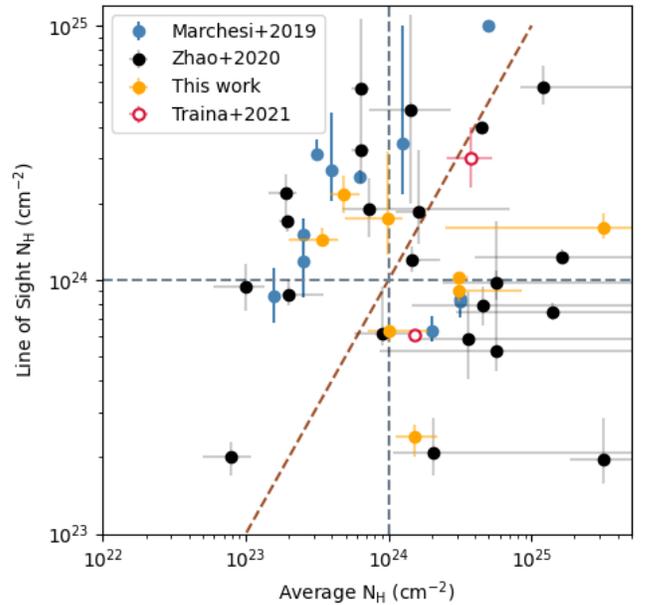

**Figure 4.** Line-of-sight hydrogen column density as a function of the average torus column density for the sources analyzed in this work (orange), in Marchesi et al. (2019) (blue), in Zhao et al. (2020b) (black) and in Traina et al. in prep. (open symbols). All plotted results correspond to the best-fit borus02 model. The dashed vertical and horizontal lines mark the CT limit, while the diagonal line is a 1:1 relation between $N_{\rm H,los}$ and $N_{\rm H,av}$.

The mentioned, phenomenological models, which do not take into account the decoupling between $N_{\rm H,los}$ and $N_{\rm H,av}$, are typically used in population synthesis models. The spectral shape of a source that is Compton-thin in the line-of-sight but has a large $N_{\rm H,av}$ might have a spectrum not so dissimilar from a source with a homogeneous Compton-thick torus, in which reflection is not taken into account in a self-consistent way. These sources, while not CT-AGN in the line-of-sight, might still contribute to the CXB in a relevant way. Therefore, we caution that our observed $20 \pm 5\%$ fraction of CT-AGN at z<0.01 should be compared with the predictions of populations synthesis models in a careful way.

### 5.2. *Clumpy torus scenario*

Based on our results, all eight sources except for ESO 323−G023 are incompatible, at 90% significance, with having the same line-of-sight and average torus column densities. Even for ESO 323−G023, the two values are compatible only at the very limits of their error range.

In Fig. 4 we plot the sources analysed by our group, comparing their line-of-sight and average torus column densities. Originally, Marchesi et al. (2018) analysed the 38 sources in their sample using borus02 with the inclination angle frozen to $\theta_i = 90º$ to obtain an estimate of $N_{H,av}$ without leaving the parameter free to vary. Zhao et al. (2020b) reanalysed most of the sample (i.e. those sources with good-enough data quality) using the same methodology as described in this work. In Fig. 4, we plot the sources as reanalyzed by Zhao et al. (2020b) when available[8]. For those with insufficient data quality, we note that the determination of $N_{H,av}$ should be taken as a rough estimate. On the other hand, a comparison of $N_{H,los}$ for both analyses has shown little difference for most sources[9] Five additional sources are not included in the plot due to having $N_H < 10^{23}$ cm$^{-2}$ (2MASXJ10523297+1036205, RBS 1037, MCG-01-30-041, B2 1204+34 and ESO 244-30, Marchesi et al. 2019). These sources, all originally CT-AGN candidates, highlight even further the importance of using *NuSTAR* to estimate $N_{H,los}$.

Fig 4 shows no strong correlation between one quantity and the other, and in fact tend to fall far from the 1-to-1 relation. This is confirmed by our statistical analysis, which yields a Pearson correlation coefficient of $\rho \approx -0.02$[10]. This means that sources that are CT in the line of sight are no more likely to have a thicker torus than other sources. This result is in agreement with that found by Zhao et al. (2020b), who analysed a sample of $\sim 100$ Compton-thin AGN in the local universe which have high-quality *NuSTAR* and soft X-ray data, finding that $N_{H,av}$ is similar at different $N_{H,los}$.

Furthermore, Fig. 4 shows how most of the analysed sources have significant differences between their estimated $N_{H,los}$ and $N_{H,av}$ values, which is a strong argument in favour of a patchy torus scenario. Observations both in the X-rays and in the near-IR band have already suggested that the structure of the obscuring material surrounding the supermassive black hole is, not surprisingly, more complex than a simple, homogeneous torus. Soft X-ray monitoring of AGN has shown variability in $N_{H,los}$ (e.g. Risaliti et al. 2002; Elvis et al. 2004; Markowitz et al. 2014). Infrared observations of AGN torus emission also support this scenario (Ramos Almeida et al. 2014). Despite this vision being largely accepted, there is still a small sample of sources for which $N_H$ variability has been confirmed, and other studies challenge this scenario after finding no variability in large samples (e.g. Laha et al. 2020, found no significant $N_H$ variability in 13/21 analysed Sy2s).

In this work we also find $N_{H,los}$ variability for MRK 622 (in observations taken 16 yr apart), and we found none for NGC 6552 (in observations taken 17 yr apart). However, a complete analysis of column density variability and torus cloud distribution is out of the scope of this work, and will be reported elsewhere. In order to draw stronger conclusions, we plan on targeting promising candidates in our sample (i.e. those with large $N_{H,los}/N_{H,av}$ differences, or those with multiple observations), and analyse them with models based on clumpy torus distributions (e.g. UXClumpy; Buchner et al. 2019). A similar idea was proposed by Yaqoob et al. (2015). We leave this analysis for future work.

### 5.3. *Agreement between models*

Generally, MYTorus and borus02 are in good agreement in their parameter estimation, particularly on their qualitative description of the source. That is, the models agree on Compton-thin vs Compton-thick classifications, both for $N_{H,los}$ and $N_{H,av}$. They are also generally compatible in their photon index estimation (except for MRK 622, ESO 112−G006), agreeing within errors for most of the sources.

This is not true when using MYTorus in a coupled configuration, as it tends to present strong disagreements with the photon index estimation, as well as systematically worse fits. Note, however, that the line-of-sight column density estimation is generally in agreement with that of the other models.

MYTorus decoupled and borus02 also agree in their qualitative description of the relative inclination of the source, with the ratio between $A_{S,90}$ and $A_{S,0}$ (showing predominance for forward or backward scattering) being consistent with the inclination angle and covering factor as estimated by borus02.

A detailed comparison of the MYTorus and borus02 performances can be found in Marchesi et al. (2019).

### 6. CONCLUSIONS

In this work we have analysed 8 CT-AGN candidates with simultaneous *XMM-Newton* and *NuSTAR* data, using the torus models MYTorus and borus02. For all of them, this is the first time their *NuSTAR* data is published. Our main conclusions are as follows:

---

[8] Note that we do not plot any source twice, but rather replace those of Marchesi et al. (2019) with the Zhao et al. (2020b) determination.

[9] Three sources were re-classified from CT to C-thin, and one from C-thin to CT. They had $N_{H,los}$ estimates close to the CT threshold.

[10] $\rho \approx 1$ or $\rho \approx -1$ indicate strong linear correlation, or anticorrelation, respectively.

- Out of the 8 analysed sources, 5 are confirmed to be CT-AGN based on their *XMM-Newton* and *NuSTAR* data. This brings the total number of *NuSTAR*-confirmed CT-AGN in the BAT catalogue at $z \leq 0.05$ to 34.

- Out of the 48 CT-AGN candidates analysed as part of our project, 24 (a $\sim 50\%$) are confirmed CT-AGN with the addition of the *NuSTAR* data. This confirms the need for *NuSTAR* in order to fully ascertain the CT nature of obscured sources.

- The percentage of confirmed CT-AGN within the BAT sample at $z \leq 0.05$ is estimated to be $\sim 8\%$ (34/417). Seven additional candidates remain to be analysed, which were not included in this work due to the fact their *NuSTAR* data was not publicly available. If all sources turn out to be CT, the total fraction of $\sim 10\%$ will still be much lower than the CT fraction predicted by population synthesis models. This is likely a result of the suppression of the intrinsic CT-AGN emission even in the $>15\,\mathrm{keV}$ band, as suggested by recent infrared studies (e.g. Yan et al. 2019; Carroll et al. 2021). It is also supported by the fact that we recover a CT-AGN fraction of $20 \pm 5\%$ within z<0.01.

- Most of the sources analysed as part of our project are best fit with a line-of-sight column density, $N_{\mathrm{H,los}}$ that differs, at $\sim 90\%$ confidence, from their average torus column density, $N_{\mathrm{H,av}}$. This supports a patchy torus hypothesis.

- We find no significant correlation between the average torus column density and the line-of-sight column density of our sample. This suggests that sources that are Compton-thick in the line-of-sight are no more likely to have a thicker torus, on average, than those that are Compton-thin.

- We find that MRK 622 presented $N_{\mathrm{H,los}}$ variability between observations at different epochs (17 yr apart), between $N_{\mathrm{H,los}} \approx 24 \times 10^{22}\,\mathrm{cm}^{-2}$ to $N_{\mathrm{H,los}} \approx 49 \times 10^{22}\,\mathrm{cm}^{-2}$.

## ACKNOWLEDGMENTS

N.T.A., M.A., R.S. and X.Z. acknowledge funding from NASA under contracts 80NSSC19K0531, 80NSSC20K0045 and, 80NSSC20K834. P.B. acknowledges financial support from the Czech Science Foundation project No. 19-05599Y. M.B. acknowledges support from the YCAA Prize Postdoctoral Fellowship. We thank Poshak Gandhi and Valentina La Parola for their helpful advice. The scientific results reported in this article are based on observations made by the X-ray observatories NuSTAR and XMM-Newton, and has made use of the NASA/IPAC Extragalactic Database (NED), which is operated by the Jet Propulsion Laboratory, California Institute of Technology under contract with NASA. We acknowledge the use of the software packages XMM-SAS and HEASoft.

CT fraction in the local Universe 15Brightman, M., Baloković, M., Stern, D., et al. 2015, ApJ, 805, 41, doi: 10.1088/0004-637X/805/1/41

Buchner, J., Brightman, M., Nandra, K., Nikutta, R., & Bauer, F. E. 2019, A&A, 629, A16, doi: 10.1051/0004-6361/201834771

Buchner, J., Georgakakis, A., Nandra, K., et al. 2015, ApJ, 802, 89, doi: 10.1088/0004-637X/802/2/89

Burlon, D., Ajello, M., Greiner, J., et al. 2011, The Astrophysical Journal, 728, 58, doi: 10.1088/0004-637x/728/1/58

Carroll, C. M., Hickox, R. C., Masini, A., et al. 2021, ApJ, 908, 185, doi: 10.3847/1538-4357/abd185

Comastri, A., Setti, G., Zamorani, G., & Hasinger, G. 1995, A&A, 296, 1. https://arxiv.org/abs/astro-ph/9409067

Cusumano, G., Segreto, A., La Parola, V., & Maselli, A. 2014, in Proceedings of Swift: 10 Years of Discovery (SWIFT 10, 132

Del Moro, A., Alexander, D. M., Bauer, F. E., et al. 2016, MNRAS, 456, 2105, doi: 10.1093/mnras/stv2748

Elvis, M., Risaliti, G., Nicastro, F., et al. 2004, ApJL, 615, L25, doi: 10.1086/424380

Georgantopoulos, I., & Akylas, A. 2019, A&A, 621, A28, doi: 10.1051/0004-6361/201833038

Gilli, R., Comastri, A., & Hasinger, G. 2007, A&A, 463, 79, doi: 10.1051/0004-6361:20066334

Harrison, F. A., Craig, W. W., Christensen, F. E., et al. 2013, ApJ, 770, 103, doi: 10.1088/0004-637X/770/2/103

Iwasawa, K., Ricci, C., Privon, G. C., et al. 2020, A&A, 640, A95, doi: 10.1051/0004-6361/202038513

Jones, D. H., Read, M. A., Saunders, W., et al. 2009, MNRAS, 399, 683, doi: 10.1111/j.1365-2966.2009.15338.x

Kaastra, J. 1992, An X-Ray Spectral Code for Optically Thin Plasmas (Internal SRON-Leiden Report, updated version 2.0)

Kalberla, P. M. W., Burton, W. B., Hartmann, D., et al. 2005, A&A, 440, 775, doi: 10.1051/0004-6361:20041864

Kammoun, E. S., Miller, J. M., Koss, M., et al. 2020, ApJ, 901, 161, doi: 10.3847/1538-4357/abb29f

Koss, M. J., Romero-Cañizales, C., Baronchelli, L., et al. 2015, ApJ, 807, 149, doi: 10.1088/0004-637X/807/2/149

Koss, M. J., Glidden, A., Baloković, M., et al. 2016, ApJL, 824, L4, doi: 10.3847/2041-8205/824/1/L4

Laha, S., Markowitz, A. G., Krumpe, M., et al. 2020, ApJ, 897, 66, doi: 10.3847/1538-4357/ab92ab

Lanzuisi, G., Ranalli, P., Georgantopoulos, I., et al. 2015, A&A, 573, A137, doi: 10.1051/0004-6361/201424924

Liedahl, D. A., Osterheld, A. L., & Goldstein, W. H. 1995, ApJL, 438, L115, doi: 10.1086/187729

Lin, D., Webb, N. A., & Barret, D. 2012, ApJ, 756, 27, doi: 10.1088/0004-637X/756/1/27

Marchesi, S., Ajello, M., Comastri, A., et al. 2017a, ApJ, 836, 116, doi: 10.3847/1538-4357/836/1/116

Marchesi, S., Ajello, M., Marcotulli, L., et al. 2018, ApJ, 854, 49, doi: 10.3847/1538-4357/aaa410

Marchesi, S., Ajello, M., Zhao, X., et al. 2019, ApJ, 882, 162, doi: 10.3847/1538-4357/ab340a

Marchesi, S., Tremblay, L., Ajello, M., et al. 2017b, ApJ, 848, 53, doi: 10.3847/1538-4357/aa8ee6

Markowitz, A. G., Krumpe, M., & Nikutta, R. 2014, MNRAS, 439, 1403, doi: 10.1093/mnras/stt2492

Marshall, F. E., Boldt, E. A., Holt, S. S., et al. 1980, ApJ, 235, 4, doi: 10.1086/157601

Masini, A., Comastri, A., Baloković, M., et al. 2016, A&A, 589, A59, doi: 10.1051/0004-6361/201527689

Mewe, R., Gronenschild, E. H. B. M., & van den Oord, G. H. J. 1985, A&AS, 62, 197

Murphy, K. D., & Yaqoob, T. 2009, MNRAS, 397, 1549, doi: 10.1111/j.1365-2966.2009.15025.x

Oda, S., Tanimoto, A., Ueda, Y., et al. 2017, ApJ, 835, 179, doi: 10.3847/1538-4357/835/2/179

Oh, K., Koss, M., Markwardt, C. B., et al. 2018, ApJS, 235, 4, doi: 10.3847/1538-4365/aaa7fd

Parisi, P., Masetti, N., Rojas, A. F., et al. 2014, A&A, 561, A67, doi: 10.1051/0004-6361/201322409

Puccetti, S., Comastri, A., Fiore, F., et al. 2014, ApJ, 793, 26, doi: 10.1088/0004-637X/793/1/26

Puccetti, S., Comastri, A., Bauer, F. E., et al. 2016, A&A, 585, A157, doi: 10.1051/0004-6361/201527189

Ramos Almeida, C., Alonso-Herrero, A., Levenson, N. A., et al. 2014, MNRAS, 439, 3847, doi: 10.1093/mnras/stu235

Ricci, C., Ueda, Y., Koss, M. J., et al. 2015, ApJL, 815, L13, doi: 10.1088/2041-8205/815/1/L13

Ricci, C., Trakhtenbrot, B., Koss, M. J., et al. 2017, ApJS, 233, 17, doi: 10.3847/1538-4365/aa96ad

Risaliti, G., Elvis, M., & Nicastro, F. 2002, ApJ, 571, 234, doi: 10.1086/324146

Rivers, E., Baloković, M., Arévalo, P., et al. 2015, ApJ, 815, 55, doi: 10.1088/0004-637X/815/1/55

Stern, D., Lansbury, G. B., Assef, R. J., et al. 2014, ApJ, 794, 102, doi: 10.1088/0004-637X/794/2/102

Tanimoto, A., Ueda, Y., Odaka, H., et al. 2019, ApJ, 877, 95, doi: 10.3847/1538-4357/ab1b20

Torres-Albà, N., Iwasawa, K., Díaz-Santos, T., et al. 2018, A&A, 620, A140, doi: 10.1051/0004-6361/201834105

Turner, T. J., Reeves, J. N., Braito, V., et al. 2020, MNRAS, 498, 1983, doi: 10.1093/mnras/staa2401

Ueda, Y., Akiyama, M., Hasinger, G., Miyaji, T., & Watson, M. G. 2014, ApJ, 786, 104, doi: 10.1088/0004-637X/786/2/104

# APPENDIX

## A. X-RAY FITTING RESULTS

**Table 5.** X-ray fitting results of MCG 07-03-007

| Model | MYTorus (Coupled) | MYTorus (Decoupled) | MYTorus (Decoupled free) | borus |
|---|---|---|---|---|
| red $\chi^2$ | 1.06 | 1.04 | 1.05 | 1.05 |
| $\chi^2$/d.o.f. | 239.45/227 | 237.17/227 | 236.25/225 | 236.98/225 |
| kT | $0.29^{+0.25}_{-0.06}$ | $0.27^{+0.15}_{-0.07}$ | $0.27^{+0.11}_{-0.07}$ | $0.29^{+0.22}_{-0.10}$ |
| $\Gamma$ | $1.78^{+0.06}_{-0.07}$ | $1.74^{+0.07}_{-0.08}$ | $1.72^{+0.17}_{-0.18}$ | $1.84^{+0.12}_{-0.15}$ |
| $N_{H,los}$ | $0.91^{+1.77}_{-0.38}$ | $0.84^{+0.07}_{-0.06}$ | $0.84^{+0.11}_{-0.11}$ | $0.90^{+0.07}_{-0.08}$ |
| $N_{H,eq}$ | $2.31^{+1.45}_{-0.95}$ | – | – | – |
| $N_{H,av}$ | – | $2.34^{+0.84}_{-0.55}$ | $2.37^{+1.77}_{-0.78}$ | $3.15^{+5.55}_{-0.28}$ |
| $A_{S90}$ | – | 1* | $1.37^{+3.85}_{-}$ | – |
| $A_{S0}$ | – | 1* | $0.80^{+0.38}_{-0.30}$ | – |
| CF (Tor) | – | – | – | $0.60^{+0.36}_{-0.10}$ |
| Cos ($\theta_{Obs}$) | $0.46^{+0.03}_{-0.11}$ | – | – | $0.57^{+0.13}_{-0.17}$ |
| $F_s$ ($10^{-3}$) | $1.85^{+5.25}_{-3.57}$ | $3.84^{+3.43}_{-2.34}$ | $3.08^{+5.17}_{-2.26}$ | $2.24^{+3.79}_{-1.50}$ |
| Norm ($10^{-4}$) | $7.99^{+2.04}_{-1.69}$ | $10.3^{+3.1}_{-2.9}$ | $17.3^{+5.3}_{-2.5}$ | $8.37^{+1.90}_{-1.56}$ |
| EW [keV] | $0.47^{+0.08}_{-0.08}$ | $0.48^{+0.17}_{-0.11}$ | $0.46^{+0.35}_{-0.15}$ | – |
| Flux (2−10 keV) [$10^{-13}$] | $4.74^{+0.29}_{-0.30}$ | $4.68^{+0.29}_{-0.29}$ | $4.70^{+0.29}_{-0.29}$ | $4.78^{+0.29}_{-0.29}$ |
| Flux (10−40 keV) [$10^{-12}$] | $5.43^{+0.17}_{-0.17}$ | $5.51^{+0.18}_{-0.18}$ | $5.46^{+0.17}_{-0.17}$ | $5.45^{+0.17}_{-0.17}$ |
| $L_{intr}$ (2−10 keV) [$10^{43}$] | $1.29^{+0.19}_{-0.19}$ | $0.89^{+0.13}_{-0.13}$ | $1.05^{+0.15}_{-0.15}$ | $1.31^{+0.18}_{-0.18}$ |
| $L_{intr}$ (15−55 keV) [$10^{43}$] | $1.73^{+0.09}_{-0.10}$ | $1.32^{+0.07}_{-0.07}$ | $1.32^{+0.07}_{-0.07}$ | $1.43^{+0.07}_{-0.07}$ |
| counts | 5235 | | | |

**Notes:** Same as Table 2.

18 Torres-Albà et al.Table 6. X-ray fitting results of ESO 426-G002

| Model | MYTorus (Coupled) | MYTorus (Decoupled) | MYTorus (Decoupled free) | borus02 |
|---|---|---|---|---|
| red $\chi^2$ | 1.20 | 1.22 | 1.11 | 1.12 |
| $\chi^2$/d.o.f. | 409.12/341 | 414.95/341 | 366.53/339 | 380.25/339 |
| $kT$ | $0.63^{+0.05}_{-0.06}$ | $0.64^{+0.05}_{-0.06}$ | $0.65^{+0.08}_{-0.07}$ | $0.64^{+0.06}_{-0.06}$ |
| $\Gamma$ | $1.59^{+0.06}_{-0.08}$ | $1.70^{+0.05}_{-0.06}$ | $2.19^{+0.13}_{-0.13}$ | $2.08^{+0.02}_{-0.03}$ |
| $N_{H,los}$ | $1.06^{+1.74}_{-0.11}$ | $0.91^{+0.05}_{-0.05}$ | $1.01^{+0.08}_{-0.08}$ | $1.02^{+0.03}_{-0.03}$ |
| $N_{H,eq}$ | $1.31^{+2.14}_{-0.14}$ | – | – | – |
| $N_{H,av}$ | – | $3.91^{+0.79}_{-0.93}$ | $3.80^{+0.84}_{-0.60}$ | $3.16^{+0.55}_{-0.30}$ |
| $A_{S90}$ | – | 1* | $4.95^{+3.29}_{-2.33}$ | – |
| $A_{S0}$ | – | 1* | $0.24^{+0.14}_{-0.14}$ | – |
| $C_{\rm F}$ | – | – | – | $0.97^{+0.02}_{-0.03}$ |
| $\cos(\theta_i)$ | $0.29^{+0.19}_{-0.08}$ | – | – | $0.87^{+0.02}_{-0.01}$ |
| $F_s$ ($10^{-3}$) | $4.21^{+2.26}_{-1.55}$ | $4.11^{+3.06}_{-1.18}$ | $1.48^{+2.03}_{-0.75}$ | $1.78^{+0.81}_{-0.25}$ |
| Norm ($10^{-3}$) | $1.99^{+0.49}_{-0.70}$ | $1.82^{+0.28}_{-0.33}$ | $8.02^{+1.94}_{-1.71}$ | $6.12^{+0.15}_{-0.83}$ |
| EW [keV] | $0.27^{+0.05}_{-0.05}$ | $0.24^{+0.04}_{-0.04}$ | $0.20^{+0.10}_{-0.10}$ | – |
| Flux (2−10 keV) [$10^{-13}$] | $5.48^{+0.26}_{-0.26}$ | $5.39^{+0.25}_{-0.25}$ | $5.39^{+0.25}_{-0.25}$ | $5.33^{+0.25}_{-0.25}$ |
| Flux (10−40 keV) [$10^{-12}$] | $8.32^{+0.22}_{-0.21}$ | $8.76^{+0.23}_{-0.23}$ | $8.49^{+0.22}_{-0.22}$ | $8.41^{+0.22}_{-0.22}$ |
| $L_{intr}$ (2-10 keV) [$10^{43}$] | $1.03^{+0.15}_{-0.15}$ | $0.78^{+0.14}_{-0.14}$ | $1.68^{+0.22}_{-0.22}$ | $1.50^{+0.15}_{-0.15}$ |
| $L_{intr}$ (15-55 keV) [$10^{43}$] | $1.84^{+0.75}_{-0.75}$ | $1.20^{+0.54}_{-0.54}$ | $0.99^{+0.42}_{-0.42}$ | $1.01^{+0.42}_{-0.42}$ |
| counts | 9492 | | | |

**Notes:** Same as Table 2.

Table 7. X-ray fitting results of LEDA 478026

| Model | MYTorus (Coupled) | MYTorus (Decoupled) | MYTorus (Decoupled free) | borus |
|---|---|---|---|---|
| red $\chi^2$ | 0.90 | 0.98 | 0.89 | 0.90 |
| $\chi^2$/d.o.f. | 134.35/150 | 146.69/150 | 131.00/148 | 133.19/148 |
| kT | $0.60^{+0.10}_{-0.14}$ | $0.61^{+0.09}_{-0.12}$ | $0.61^{+0.13}_{-0.15}$ | $0.61^{+0.11}_{-0.15}$ |
| $\Gamma$ | $1.62^{+0.06}_{-0.07}$ | $1.59^{+0.06}_{-0.08}$ | $1.80^{+0.06}_{-0.08}$ | $1.72^{+0.07}_{-0.09}$ |
| $N_{H,los}$ | $1.04^{+0.88}_{-0.67}$ | $0.89^{+0.10}_{-0.09}$ | $1.45^{+0.13}_{-0.17}$ | $1.44^{+0.16}_{-0.09}$ |
| $N_{H,eq}$ | $1.28^{+0.73}_{-0.20}$ | – | – | – |
| $N_{H,av}$ | – | $2.56^{+1.00}_{-0.66}$ | $0.33^{+0.67}_{-0.13}$ | $0.34^{+0.11}_{-0.14}$ |
| $A_{S90}$ | – | 1* | $0.33^{+0.14}_{-}$ | – |
| $A_{S0}$ | – | 1* | $0.04^{+0.25}_{-}$ | – |
| CF (Tor) | – | – | – | $0.15^{+0.05}_{-}$ |
| Cos ($\theta_{Obs}$) | $0.29^{+0.18}_{-0.14}$ | – | – | $0.05^{+0.23}_{-}$ |
| $F_s$ ($10^{-3}$) | $7.85^{+7.16}_{-4.50}$ | $1.25^{+1.15}_{-0.56}$ | $2.63^{+2.85}_{-1.65}$ | $3.22^{+1.33}_{-4.55}$ |
| Norm ($10^{-3}$) | $1.00^{+0.23}_{-0.21}$ | $0.573^{+0.110}_{-0.122}$ | $3.35^{+0.44}_{-0.77}$ | $2.49^{+0.53}_{-0.09}$ |
| EW [keV] | $0.40^{+0.11}_{-0.11}$ | $0.36^{+0.11}_{-0.12}$ | $0.39^{+0.06}_{-0.17}$ | – |
| Flux (2−10 keV) [$10^{-13}$] | $2.70^{+0.18}_{-0.18}$ | $2.65^{+0.18}_{-0.17}$ | $2.68^{+0.18}_{-0.18}$ | $2.69^{+0.18}_{-0.18}$ |
| Flux (10−40 keV) [$10^{-12}$] | $3.68^{+0.16}_{-0.16}$ | $3.83^{+0.16}_{-0.16}$ | $3.60^{+0.15}_{-0.15}$ | $3.60^{+0.15}_{-0.15}$ |
| $L_{intr}$ (2-10 keV) [$10^{43}$] | $1.63^{+0.41}_{-0.41}$ | $1.00^{+0.13}_{-0.13}$ | $4.21^{+0.52}_{-0.52}$ | $3.61^{+1.06}_{-1.05}$ |
| $L_{intr}$ (15-55 keV) [$10^{43}$] | $2.80^{+0.19}_{-0.19}$ | $1.81^{+0.14}_{-0.14}$ | $4.90^{+0.26}_{-0.27}$ | $4.73^{+0.26}_{-0.26}$ |
| counts | 3761 | | | |

**Notes:** Same as Table 2.

Table 8. X-ray fitting results of MRK 622

| Model | MYTorus (Coupled) | MYTorus (Decoupled) | MYTorus (Decoupled free) | borus02 |
|---|---|---|---|---|
| red $\chi^2$ | 1.26 | 1.25 | 1.26 | 1.22 |
| $\chi^2$/d.o.f. | 232.20/184 | 231.60/185 | 231.43/183 | 224.14/183 |
| $kT$ | $0.66^{+0.10}_{-0.08}$ | $0.67^{+0.10}_{-0.08}$ | $0.67^{+0.10}_{-0.08}$ | $0.64^{+0.09}_{-0.09}$ |
| $\Gamma$ | $1.50^{+0.11}_{-}$ | $1.54^{+0.12}_{-}$ | $1.54^{+0.14}_{-}$ | $1.74^{+0.17}_{-0.19}$ |
| $N_{H,los}$ | $0.39^{+0.49}_{-0.22}$ | $0.19^{+0.04}_{-0.04}$ | $0.19^{+0.04}_{-0.04}$ | $0.23^{+0.06}_{-0.05}$ |
| $N_{H,eq}$ | $1.97^{+2.48}_{-1.14}$ | − | − | − |
| $N_{H,av}$ | − | $1.68^{+1.43}_{-0.98}$ | $1.29^{+2.80}_{-0.66}$ | $1.55^{+0.62}_{-0.67}$ |
| $A_{S90}$ | − | 1* | $1.88^{+3.38}_{-}$ | − |
| $A_{S0}$ | − | 1* | $0.51^{+1.51}_{-}$ | − |
| $C_{\rm F}$ | − | − | − | $1.00^{-}_{-0.40}$ |
| $\cos(\theta_i)$ | $0.49^{+0.00}_{-0.02}$ | − | − | $0.8^{-}_{-}$ |
| $F_s$ ($10^{-2}$) | $1.36^{+1.32}_{-1.11}$ | $1.38^{+1.90}_{-}$ | $1.42^{+1.89}_{-}$ | $1.34^{+1.00}_{-0.94}$ |
| Norm ($10^{-4}$) | $2.33^{+0.84}_{-0.56}$ | $2.37^{+0.95}_{-0.79}$ | $2.36^{+1.09}_{-0.79}$ | $4.02^{+3.62}_{-1.80}$ |
| EW [keV] | $0.12^{+0.08}_{-0.08}$ | $0.13^{+0.08}_{-0.08}$ | $0.12^{+0.17}_{-}$ | ... |
| Flux (2−10 keV) [$10^{-13}$] | $4.76^{+0.42}_{-0.42}$ | $4.98^{+0.44}_{-0.44}$ | $4.99^{+0.44}_{-0.44}$ | $4.94^{+0.43}_{-0.43}$ |
| Flux (10−40 keV) [$10^{-12}$] | $2.52^{+0.10}_{-0.10}$ | $2.62^{+0.10}_{-0.10}$ | $2.59^{+0.10}_{-0.10}$ | $2.50^{+0.10}_{-0.10}$ |
| $L_{intr}$ (2−10 keV) [$10^{42}$] | $1.34^{+0.19}_{-0.20}$ | $1.25^{+0.19}_{-0.19}$ | $1.27^{+0.19}_{-0.19}$ | $1.70^{+0.23}_{-0.23}$ |
| $L_{intr}$ (15−55 keV) [$10^{42}$] | $2.68^{+0.12}_{-0.12}$ | $2.77^{+0.14}_{-0.14}$ | $2.79^{+0.13}_{-0.14}$ | $2.21^{+0.10}_{-0.10}$ |
| counts | 4938 | | | |

**Notes:** Same as Table 2.

Table 9. X-ray fitting results of Mrk 622 - with archival data

| Model | MYTorus (Coupled) | MYTorus (Decoupled) | MYTorus (Decoupled free) | borus02 |
|---|---|---|---|---|
| red $\chi^2$ | 1.38 | 1.23 | 1.25 | 1.20 |
| $\chi^2$/d.o.f. | 279.01/200 | 245.62/200 | 245.48/198 | 237.24/198 |
| $kT$ | $0.66^{+0.07}_{-0.06}$ | $0.65^{+0.07}_{-0.07}$ | $0.65^{+0.07}_{-0.07}$ | $0.64^{+0.07}_{-0.07}$ |
| $\Gamma$ | $1.48^{+0.07}_{-}$ | $1.55^{+0.10}_{-0.13}$ | $1.53^{+0.16}_{-}$ | $1.74^{+0.12}_{-0.13}$ |
| $N_{H,los}$ | $0.24^{+0.09}_{-0.13}$ | $0.20^{+0.03}_{-0.03}$ | $0.20^{+0.04}_{-0.04}$ | $0.24^{+0.03}_{-0.04}$ |
| $N_{H,eq}$ | $1.20^{+0.46}_{-0.63}$ | − | − | − |
| $N_{H,av}$ | − | $1.67^{+1.29}_{-0.84}$ | $1.99^{+1.66}_{-1.33}$ | $1.50^{+0.65}_{-0.38}$ |
| $A_{S90}$ | − | 1* | $0.29^{+3.89}_{-}$ | − |
| $A_{S0}$ | − | 1* | $1.07^{+0.91}_{-}$ | − |
| $C_{\rm F}$ | − | − | − | $1.00^{-}_{-0.40}$ |
| $\cos(\theta_i)$ | $0.49^{+0.01}_{-0.01}$ | − | − | $0.84^{-}_{-}$ |
| $F_{\rm s}$ $(10^{-2})$ | $2.19^{+1.61}_{-1.11}$ | $2.06^{+1.74}_{-1.13}$ | $2.29^{+1.66}_{-0.93}$ | $1.76^{+1.16}_{-0.80}$ |
| Norm $(10^{-4})$ | $2.32^{+1.66}_{-1.03}$ | $2.41^{+0.75}_{-0.73}$ | $2.38^{+0.86}_{-0.71}$ | $4.10^{+1.89}_{-1.34}$ |
| $C$ | $0.58^{+0.07}_{-0.07}$ | $0.98^{+0.19}_{-0.16}$ | $0.99^{+0.20}_{-0.16}$ | $0.95^{+0.18}_{-0.16}$ |
| $N_{\rm H,los,2}$ | $=N_{H,los}$ | $0.51^{+0.17}_{-0.12}$ | $0.51^{+0.18}_{-0.12}$ | $0.49^{+0.12}_{-0.10}$ |
| counts | 6849 | | | |

**Notes:** Same as Table 2, plus:

C: Cross-normalization constant between observations.

$N_{H,los,2}$: Line-of-sight torus hydrogen column density of the archived observation, in units of $10^{24}$ cm$^{-2}$.

Table 10. X-ray fitting results of NGC 6552

| Model | MYTorus (Coupled) | MYTorus (Decoupled) | MYTorus (Decoupled free) | borus02 |
|---|---|---|---|---|
| red $\chi^2$ | 1.47 | 1.12 | 1.10 | 1.12 |
| $\chi^2$/d.o.f. | 250.20/170 | 190.19/170 | 183.78/168 | 187.24/168 |
| $kT$ | $0.65^{+0.06}_{-0.06}$ | $0.64^{+0.07}_{-0.07}$ | $0.66^{+0.11}_{-0.09}$ | $0.65^{+0.09}_{-0.09}$ |
| $\Gamma$ | $2.43^{+0.04}_{-0.05}$ | $1.62^{+0.10}_{-0.09}$ | $1.99^{+0.12}_{-0.15}$ | $1.84^{+0.11}_{-0.11}$ |
| $N_{H,los}$ | $1.39^{+0.23}_{-0.09}$ | $1.64^{+0.27}_{-0.22}$ | $2.50^{+0.66}_{-0.35}$ | $2.27^{+0.40}_{-0.37}$ |
| $N_{H,eq}$ | $7.01^{+1.11}_{-0.44}$ | – | – | – |
| $N_{H,av}$ | – | $0.38^{+0.11}_{-0.08}$ | $0.38^{+0.21}_{-0.07}$ | $0.42^{+0.13}_{-0.08}$ |
| $A_{S90}$ | – | $1^*$ | $0.69^{+0.21}_{-0.18}$ | – |
| $A_{S0}$ | – | $1^*$ | $0.09^{+0.26}_{-}$ | – |
| $C_{\rm F}$ (Tor) | – | – | – | $0.34^{+0.06}_{-0.04}$ |
| $\cos(\theta_i)$ | $0.49^{+0.01}_{-0.01}$ | – | – | $0.25^{+0.11}_{-0.08}$ |
| $F_s$ ($10^{-4}$) | $0^*$ | $49.9^{+16.0}_{-18.7}$ | $9.35^{+2.35}_{-0.60}$ | $16.8^{+50.8}_{-11.9}$ |
| Norm ($10^{-2}$) | $1.53^{+0.27}_{-0.19}$ | $0.180^{+0.026}_{-0.035}$ | $1.44^{+0.38}_{-0.49}$ | $0.684^{+0.171}_{-0.245}$ |
| EW [keV] | $0.48^{+0.06}_{-0.06}$ | $0.46^{+0.05}_{-0.05}$ | $0.41^{+0.10}_{-0.10}$ | – |
| Flux (2−10 keV) [$10^{-13}$] | $6.29^{+0.35}_{-0.35}$ | $6.55^{+0.36}_{-0.36}$ | $6.44^{+0.36}_{-0.36}$ | $6.51^{+0.36}_{-0.36}$ |
| Flux (10−40 keV) [$10^{-12}$] | $4.56^{+0.15}_{-0.15}$ | $4.95^{+0.16}_{-0.16}$ | $4.93^{+0.16}_{-0.16}$ | $4.92^{+0.16}_{-0.16}$ |
| $L_{intr}$ (2-10 keV) [$10^{43}$] | $4.09^{+0.87}_{-0.87}$ | $1.23^{+0.24}_{-0.24}$ | $5.60^{+0.83}_{-0.83}$ | $3.46^{+0.54}_{-0.54}$ |
| $L_{intr}$ (15-55 keV) [$10^{43}$] | $1.20^{+0.08}_{-0.07}$ | $2.11^{+0.19}_{-0.19}$ | $4.04^{+0.40}_{-0.40}$ | $3.55^{+0.32}_{-0.32}$ |
| counts | 6305 | | | |

**Notes:** Same as Table 2.

Table 11. X-ray fitting results of NGC 6552 - with archival data

| Model | MYTorus (Coupled) | MYTorus (Decoupled) | MYTorus (Decoupled free) | borus02 |
|---|---|---|---|---|
| red $\chi^2$ | 1.49 | 1.16 | 1.17 | 1.17 |
| $\chi^2$/d.o.f. | 288.66/194 | 225.69/194 | 223.85/192 | 225.31/192 |
| $kT$ | $0.66^{+0.05}_{-0.05}$ | $0.66^{+0.08}_{-0.07}$ | $0.65^{+0.07}_{-0.07}$ | $0.65^{+0.06}_{-0.07}$ |
| $\Gamma$ | $2.39^{+0.07}_{-0.05}$ | $1.66^{+0.09}_{-0.09}$ | $1.75^{+0.11}_{-0.14}$ | $1.76^{+0.08}_{-0.12}$ |
| $N_{H,los}$ | $1.78^{+0.18}_{-0.17}$ | $1.66^{+0.29}_{-0.24}$ | $2.20^{+1.15}_{-0.42}$ | $2.18^{+0.38}_{-0.35}$ |
| $N_{H,eq}$ | $8.96^{+0.91}_{-0.85}$ | – | – | – |
| $N_{H,av}$ | – | $0.42^{+0.13}_{-0.08}$ | $0.46^{+0.40}_{-0.14}$ | $0.48^{+0.15}_{-0.13}$ |
| $A_{S90}$ | – | 1* | $0.82^{+0.39}_{-0.25}$ | – |
| $A_{S0}$ | – | 1* | $0.43^{+0.28}_{-0.36}$ | – |
| $C_F$ | – | – | – | $0.40^{+0.09}_{-0.05}$ |
| $\cos(\theta_i)$ | $0.49^{+0.01}_{-0.01}$ | – | – | $0.34^{+0.11}_{-0.11}$ |
| $F_s$ ($10^{-3}$) | $0.121^{+0.354}_{-}$ | $6.07^{+1.86}_{-1.53}$ | $2.73^{+6.71}_{-2.27}$ | $5.72^{+2.53}_{-3.87}$ |
| Norm ($10^{-3}$) | $15.2^{+2.3}_{-1.9}$ | $1.71^{+0.30}_{-0.25}$ | $4.33^{+1.20}_{-1.09}$ | $4.51^{+1.05}_{-1.50}$ |
| $C$ | $1.00^{+0.10}_{-0.09}$ | $0.96^{+0.09}_{-0.08}$ | $0.96^{+0.09}_{-0.09}$ | $0.96^{+0.09}_{-0.09}$ |
| $N_{H,los,2}$ | $=N_{H,los}$ | $=N_{H,los}$ | $=N_{H,los}$ | $=N_{H,los}$ |
| counts | 6849 | | | |

**Notes:** Same as Table 9.

Table 12. X-ray fitting results of ESO 323-G023

| Model | MYTorus (Coupled) | MYTorus (Decoupled) | MYTorus (Decoupled free) | borus |
|---|---|---|---|---|
| red $\chi^2$ | 1.10 | 0.97 | 0.98 | 0.97 |
| $\chi^2$/d.o.f. | 132.21/122 | 118.58/122 | 117.09/120 | 116.9/120 |
| kT | $0.52^{+0.12}_{-0.19}$ | $0.32^{+0.17}_{-0.06}$ | $0.31^{+0.16}_{-0.06}$ | $0.30^{+0.12}_{-0.06}$ |
| $\Gamma$ | $2.53^{-}_{-0.13}$ | $1.91^{+0.11}_{-0.13}$ | $1.96^{+0.61}_{-0.40}$ | $2.02^{+0.13}_{-0.30}$ |
| $N_{H,los}$ | $1.37^{+0.69}_{-0.49}$ | $1.45^{+0.63}_{-0.33}$ | $1.93^{-}_{-0.85}$ | $1.75^{+1.46}_{-0.49}$ |
| $N_{H,eq}$ | $1.75^{+0.63}_{-0.21}$ | − | − | − |
| $N_{H,av}$ | − | $1.00^{+0.32}_{-0.21}$ | $0.83^{+0.96}_{-0.34}$ | $0.98^{+0.28}_{-0.49}$ |
| $A_{S90}$ | − | 1* | $1.57^{+9.29}_{-1.25}$ | − |
| $A_{S0}$ | − | 1* | $0.40^{+1.54}_{-}$ | − |
| CF (Tor) | − | − | − | $0.61^{+0.37}_{-0.06}$ |
| Cos ($\theta_{Obs}$) | $0.31^{+0.10}_{-0.06}$ | − | − | $0.55^{+0.27}_{-}$ |
| $F_s$ ($10^{-4}$) | $9.26^{+19.82}_{-4.15}$ | $75.0^{+7.7}_{-4.6}$ | $55.5^{+11275}_{-4.52}$ | $47.5^{+1.1}_{-2.9}$ |
| Norm ($10^{-3}$) | $7.70^{+2.25}_{-3.65}$ | $1.00^{+0.22}_{-0.22}$ | $1.56^{+0.40}_{-0.37}$ | $1.82^{+0.43}_{-0.41}$ |
| EW [keV] | $0.76^{+0.15}_{-0.15}$ | $1.14^{+0.35}_{-0.39}$ | $1.06^{+0.37}_{-0.37}$ | − |
| Flux (2−10 keV) [$10^{-13}$] | $1.99^{+0.14}_{-0.14}$ | $2.02^{+0.14}_{-0.14}$ | $2.01^{+0.14}_{-0.14}$ | $2.01^{+0.14}_{-0.14}$ |
| Flux (10−40 keV) [$10^{-12}$] | $1.44^{+0.09}_{-0.09}$ | $1.65^{+0.10}_{-0.11}$ | $1.65^{+0.11}_{-0.11}$ | $1.61^{+0.10}_{-0.10}$ |
| $L_{intr}$ (2-10 keV) [$10^{42}$] | $5.11^{+1.02}_{-1.04}$ | $1.61^{+0.25}_{-0.25}$ | $2.31^{+0.33}_{-0.33}$ | $2.60^{+0.35}_{-0.36}$ |
| $L_{intr}$ (15-55 keV) [$10^{42}$] | $1.49^{+0.19}_{-0.18}$ | $1.56^{+0.31}_{-0.31}$ | $2.04^{+0.54}_{-0.53}$ | $2.18^{+0.47}_{-0.47}$ |
| counts | 2566 | | | |

**Notes:** Same as Table 2.

Table 13. X-ray fitting results of CGCG 475-040

| Model | MYTorus (Coupled) | MYTorus (Decoupled) | MYTorus (Decoupled free) | borus (with BAT) |
|---|---|---|---|---|
| red $\chi^2$ | 1.18 | 1.12 | 1.13 | 1.07 |
| $\chi^2$/d.o.f. | 135.05/114 | 128.24/114 | 126.14/112 | 124.87/116 |
| kT | $0.87^{+0.12}_{-0.08}$ | $0.88^{+0.11}_{-0.10}$ | $0.87^{+0.12}_{-0.09}$ | $0.88^{+0.13}_{-0.12}$ |
| $\Gamma$ | $2.03^{+0.20}_{-0.09}$ | $2.16^{+0.08}_{-0.09}$ | $2.54^{-}_{-0.39}$ | $1.72^{+0.15}_{-0.12}$ |
| $N_{H,los}$ | $1.56^{+5.50}_{-1.12}$ | $1.36^{+0.24}_{-0.16}$ | $1.19^{+0.41}_{-0.41}$ | $1.60^{+0.23}_{-0.15}$ |
| $N_{H,eq}$ | $4.58^{+4.61}_{-2.35}$ | – | – | – |
| $N_{H,av}$ | – | $4.01^{+5.21}_{-1.43}$ | $3.70^{+3.40}_{-0.98}$ | $31.6^{-}_{-29.1}$ |
| $A_{S90}$ | – | 1* | $8.56^{+11.05}_{-8.42}$ | – |
| $A_{S0}$ | – | 1* | $0.99^{+2.49}_{-0.54}$ | – |
| CF (Tor) | – | – | – | $0.90^{+0.06}_{-0.21}$ |
| Cos ($\theta_{Obs}$) | $0.47^{+0.02}_{-0.15}$ | – | – | $0.87^{+0.01}_{-0.11}$ |
| $F_s$ ($10^{-4}$) | 0* | $6.27^{+13.32}_{-}$ | 0* | $14.3^{+42.7}_{-11.1}$ |
| Norm ($10^{-3}$) | $3.06^{+4.06}_{-0.63}$ | $3.20^{+0.86}_{-0.82}$ | $5.19^{+7.43}_{-4.01}$ | $2.32^{+0.96}_{-0.31}$ |
| C | – | – | – | $2.38^{+0.49}_{-0.48}$ |
| $E_{cut}$ [keV] | – | – | – | $21.0^{+17.7}_{-}$ |
| EW [keV] | $0.87^{+0.15}_{-0.15}$ | $0.88^{+0.21}_{-0.29}$ | $0.83^{+0.28}_{-0.29}$ | – |
| Flux (2−10 keV) [$10^{-13}$] | $2.19^{+0.17}_{-0.17}$ | $2.18^{+0.17}_{-0.17}$ | $2.16^{+0.17}_{-0.17}$ | $2.19^{+0.17}_{-0.17}$ |
| Flux (10−40 keV) [$10^{-12}$] | $2.88^{+0.14}_{-0.14}$ | $2.81^{+0.13}_{-0.14}$ | $2.77^{+0.14}_{-0.14}$ | $2.54^{+0.12}_{-0.12}$ |
| $L_{intr}$ (2-10 keV) [$10^{43}$] | $1.91^{+0.91}_{-0.92}$ | $1.48^{+0.74}_{-0.74}$ | $0.68^{+0.31}_{-0.30}$ | $1.71^{+0.83}_{-0.83}$ |
| $L_{intr}$ (15-55 keV) [$10^{43}$] | $1.48^{+0.15}_{-0.15}$ | $1.04^{+0.12}_{-0.12}$ | $0.40^{+0.06}_{-0.06}$ | $0.92^{+0.12}_{-0.12}$ |
| counts | 2878 | | | |

**Notes:** Same as Table 2, plus:
C: Cross-normalization constant with BAT observation.
$E_{cut}$: Cut-off energy of the intrinsic powerlaw.

B. FIGURES

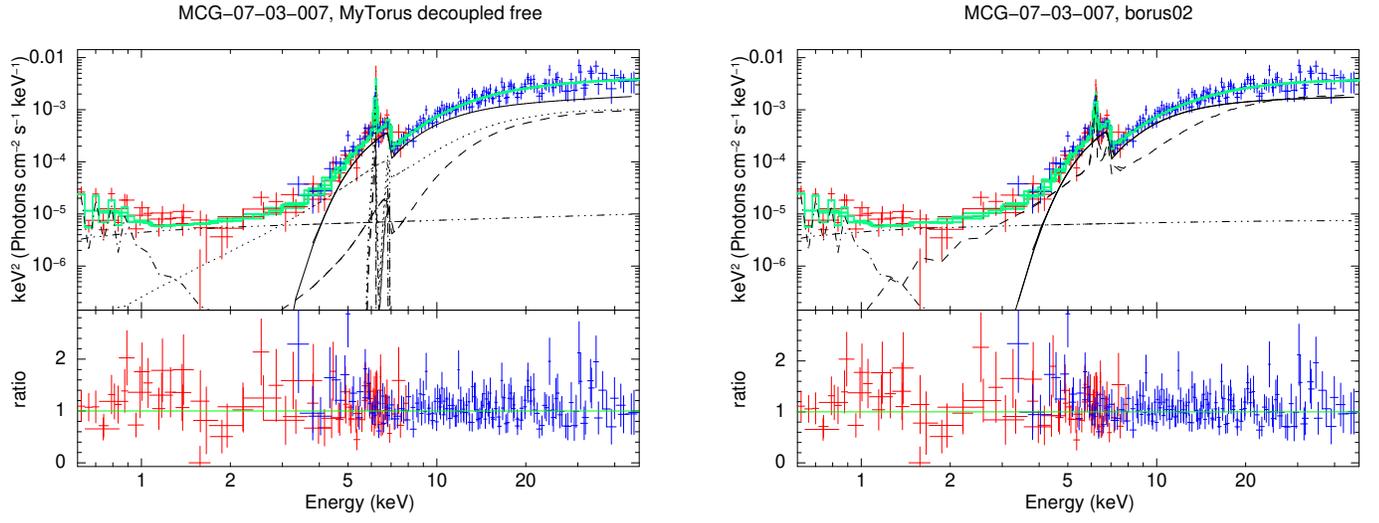

**Figure 5.** Same as Fig. 1, for MCG-07-03-007.

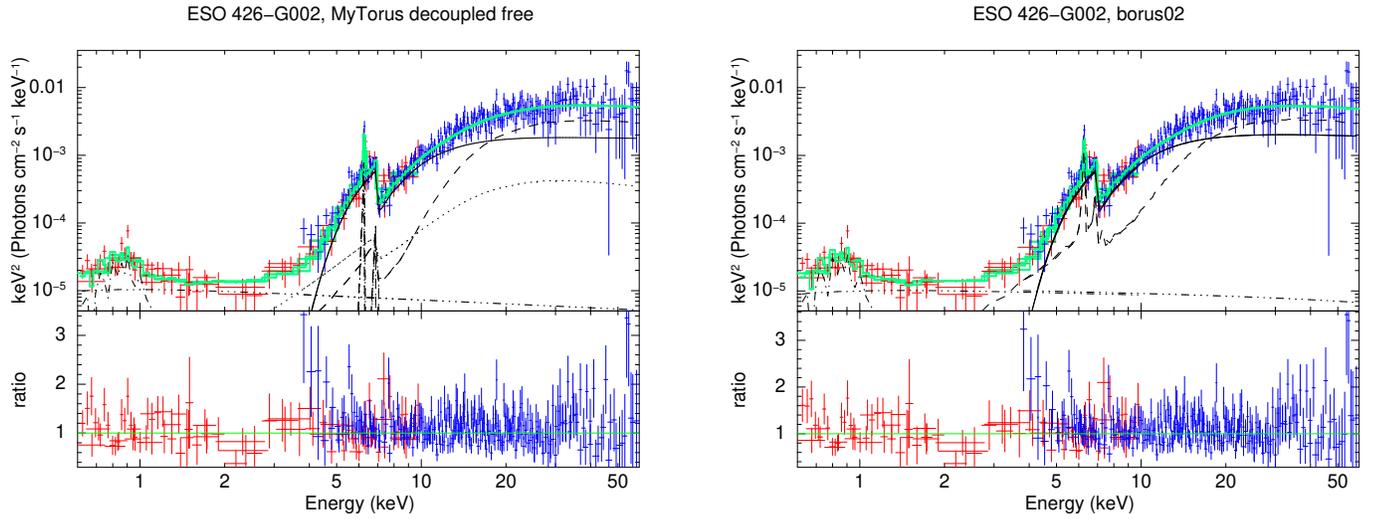

**Figure 6.** Same as Fig. 1, for ESO 426−G002.

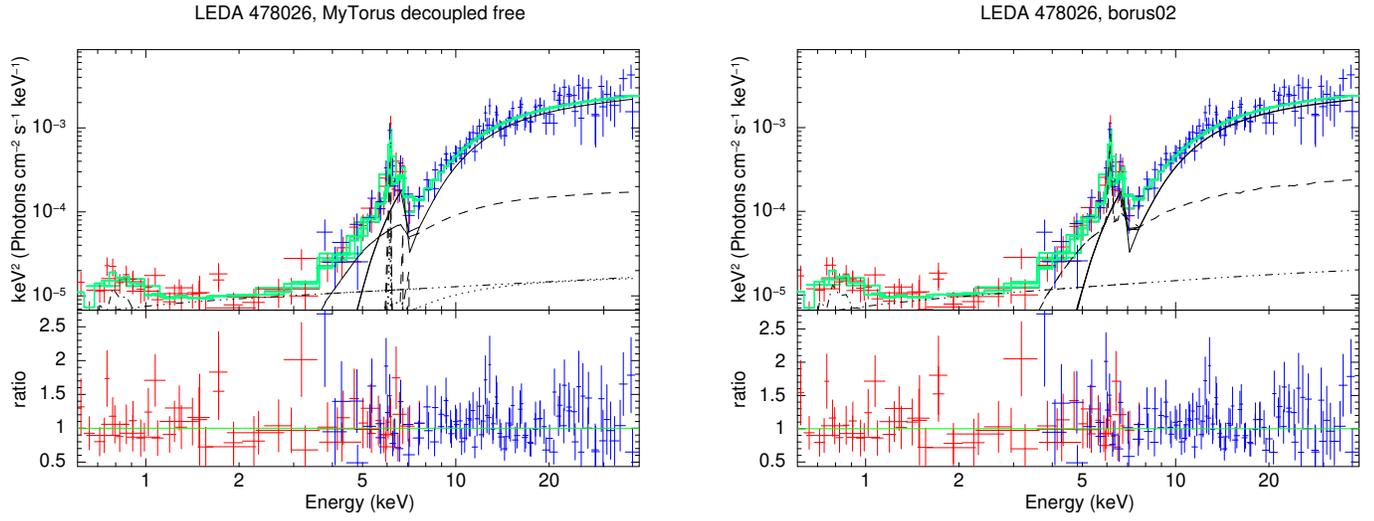

**Figure 7.** Same as Fig. 1, for LEDA 478026.

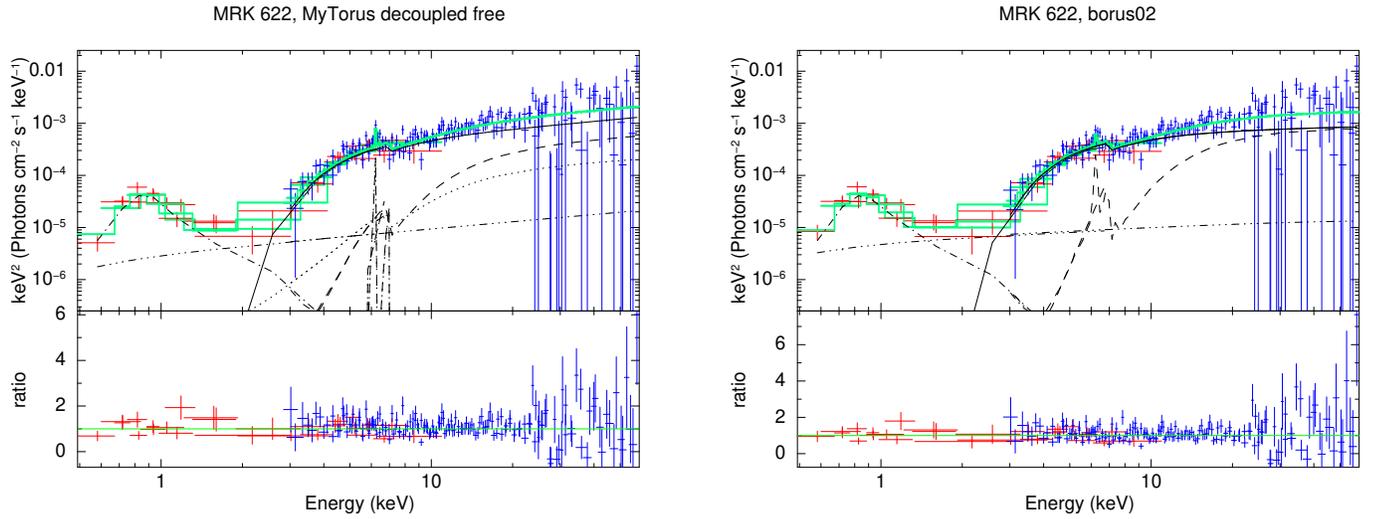

**Figure 8.** Same as Fig. 1, for MRK 622.

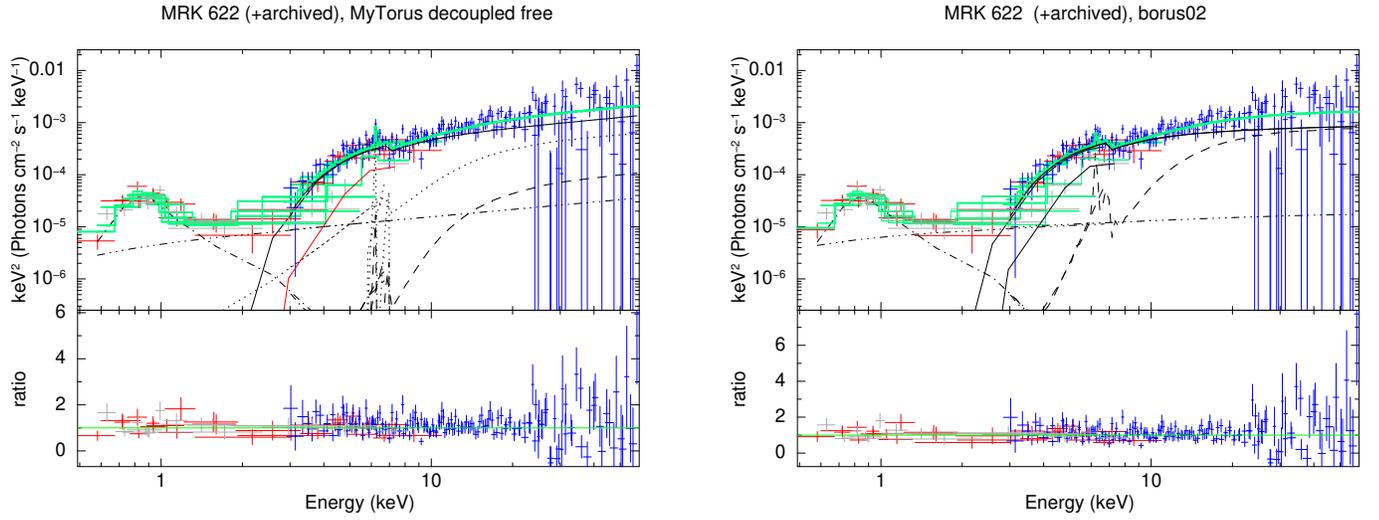

**Figure 9.** Same as Fig. 1, for MRK 622, with the inclusion of a second *XMM-Newton* observation taken from the archive, plotted in grey crosses. For this source the two *XMM-Newton* observations were fitted with different $N_{\rm H,los}$, so we add the line-of-sight component for the second observation as a solid, red curve.

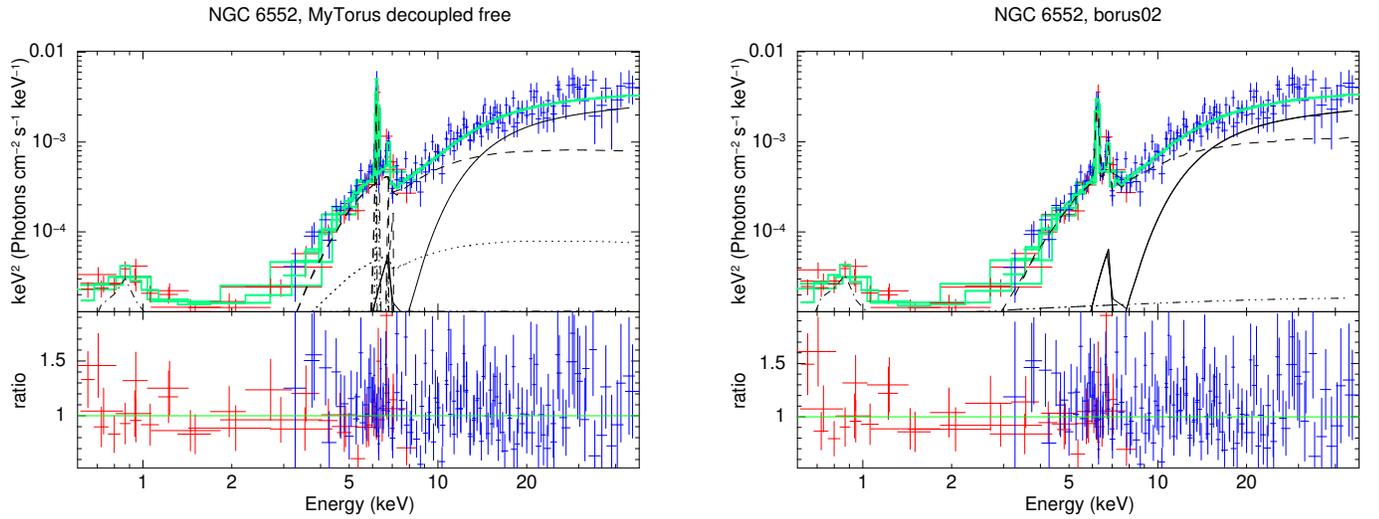

**Figure 10.** Same as Fig. 1, for NGC 6552.

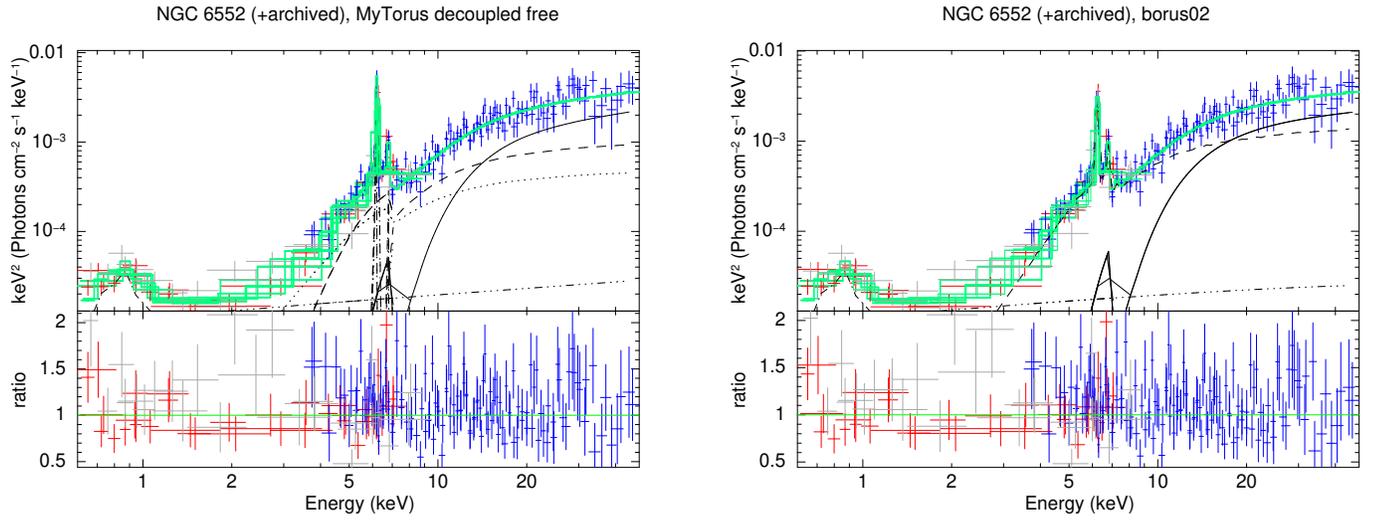

**Figure 11.** Same as Fig. 1, for NGC 6552, with the inclusion of a second *XMM-Newton* observation taken from the archive, plotted in grey crosses.

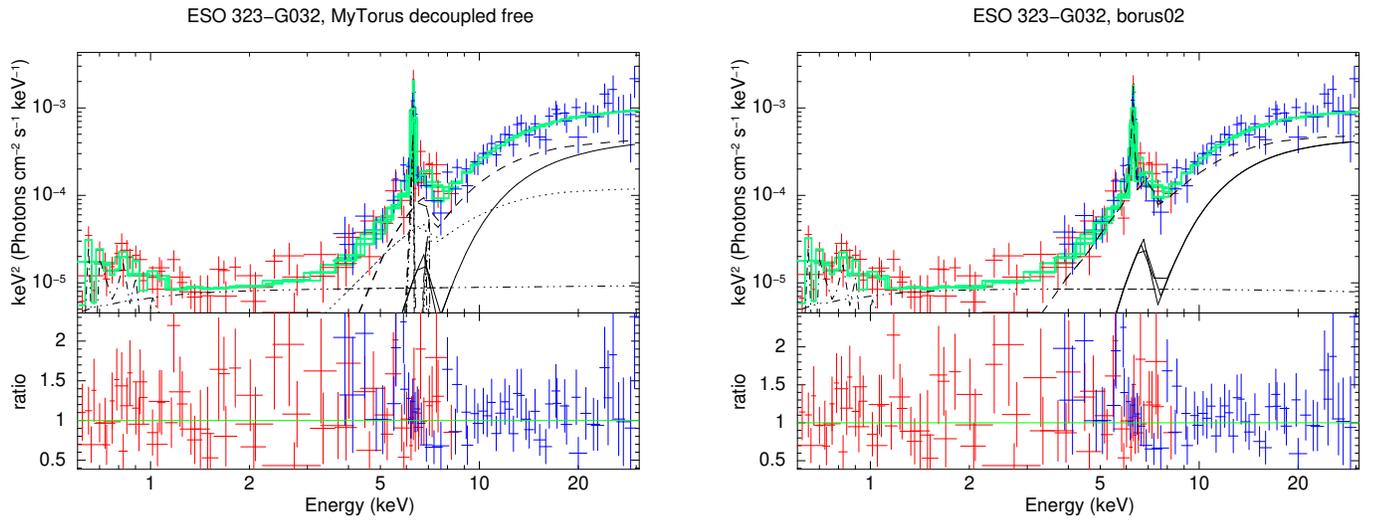

**Figure 12.** Same as Fig. 10, for ESO 323-G032.

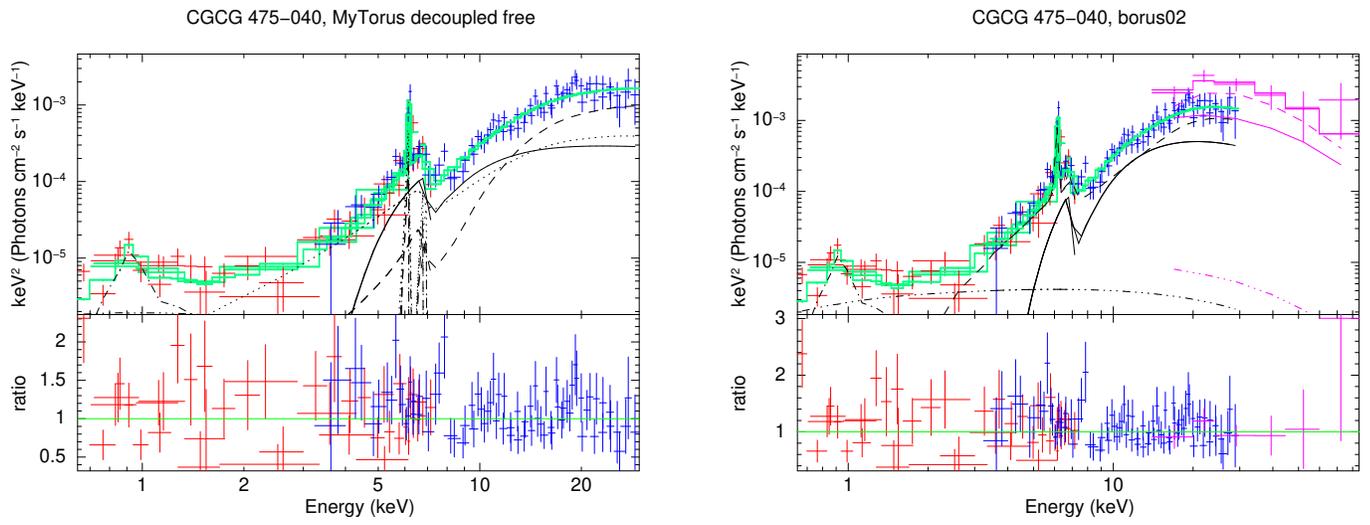

**Figure 13.** Same as Fig. 10, for CGCG 475-040.